\documentclass[11pt]{article}

\usepackage[final]{acl}

\usepackage{times}
\usepackage{latexsym}

\usepackage[T1]{fontenc}

\usepackage[utf8]{inputenc}

\usepackage{microtype}
\usepackage{inconsolata}

\usepackage{graphicx}

\usepackage{amsmath,amssymb,booktabs,multirow,subcaption}
\newcommand{\pp}{\,pp}

\title{What Counts as Evidence? Epistemic Cue Utilization in Large Language Models}

\author{
    Jinchang Zhang \\
    Indiana University Bloomington \\
    \texttt{chang2021cv@gmail.com}
    \And
    Kaijun Jiang \\
    University of Jyväskylä \\
    \texttt{kaijun.k.jiang@gmail,com}
    \And
    Guoyu Lu \\
    Indiana University Bloomington \\
    \texttt{guoyulu62@gmail.com}
}

\begin{document}

\maketitle

% =========================================================
% Abstract
% =========================================================

\begin{abstract}
Which information actually shapes a language model's judgment when sources disagree? We study epistemic cue utilization through a controlled benchmark of 500 scenarios and five complementary tracks separating cue identification, cue influence, and competition between source, temporal, and presentation cues. Across 13 LLMs, nearly identical recognition of relative source reliability can accompany effects ranging from negligible to complete choice reversal when source histories are exchanged. The resulting profiles distinguish models that rely on source history from those more responsive to evidence lineage, event time, or presentation order. We then intervene on three Qwen checkpoints using matched sham material, fact restatement, structured evidence records, and an additional weighting principle. Restatement increases reliability sensitivity in the two 27B models, and adding the principle increases it by a further 39.2 and 47.2 percentage points relative to the ledger. The 9B model instead changes its sensitivity to evidence lineage without a comparable reliability response. Verified-state accuracy remains 100\% throughout these interventions. Together, the benchmark and interventions show that identifying an epistemic cue and allowing it to influence judgment are distinct, experimentally measurable aspects of model behavior.
Code and benchmark artifacts are available at
\url{https://anonymous.4open.science/r/epistemic-benchmark/}.
\end{abstract}

% =========================================================
% 1. Introduction
% =========================================================

\section{Introduction}
\label{sec:introduction}

Large language models (LLMs) are increasingly required to make judgments in environments where information is incomplete, conflicting, and provided by multiple sources. Reliable judgment in such settings means neither simply accepting the most recent information nor blindly preserving a prior answer, but instead weighing the source of information, the source's past record, the strength and independence of the evidence, and the temporal relations among the events being described. This raises the following question: when multiple informational cues compete, which cues actually influence the model's final judgment?

Existing work on LLM belief updating, conflicting contexts, and social information use often treats whether a model changes its answer after receiving new information as the primary outcome. However, whether the answer changes conflates several fundamentally different possibilities: the model may fail to identify a relevant cue; it may identify the cue but remain behaviorally insensitive to changes in it; or it may indeed use source and evidential information while responding more strongly to other features, such as event time, repeated reports, or presentation order.
Several traditions in cognitive science provide useful functional distinctions for decomposing this problem. Research on source monitoring examines how people attribute remembered information to its origins based on perceptual and reflectively generated features \citep{johnson1993source}. Research on source credibility shows that even when informational content is comparable, its influence on judgment varies with the credibility of its source \citep{hovland1951influence,birnbaum1976combining}, while research on epistemic vigilance further argues that reasoners actively evaluate whether information and its source are worthy of trust \citep{sperber2010epistemic}. Research on advice taking further shows that decision makers can form and update reputational judgments about advisors based on their prior performance and adjust their reliance on subsequent advice accordingly \citep{yaniv2000advice}. Together, these traditions motivate separating the content of a report from judgments about its source, and treating source identity and historical reliability as distinct features of epistemic judgment rather than incidental properties of the input.

Crucially, however, being able to identify a judgment-relevant feature is not equivalent to that feature actually controlling judgment. Research on information integration and cue utilization has long examined how multiple pieces of information are assigned different functional weights in judgment, and how variations in cue and task properties shape those weights and resulting judgments \citep{brehmer1972cue,anderson1981foundations}.
From this perspective, different decision makers may have access to exactly the same information while assigning substantially different influence to it.
In addition, research on belief updating shows that even when the underlying evidence is held constant, final judgments may depend on the order in which evidence is presented, producing primacy or recency effects \citep{hogarth1992order}. These traditions therefore jointly motivate a central distinction in our evaluation: whether a cue can be identified and how strongly that cue influences a final judgment are related but separable behavioral questions.
We bring these functional distinctions to the study of cue utilization in epistemic judgment under uncertainty in LLMs. We do not assume that language models instantiate the same psychological mechanisms as humans. Instead, we use these cognitive-science distinctions to define experimentally separable behavioral measurements. We treat source identity, historical source reliability, current evidence quality, evidence independence, and event time as epistemic cues: features of information, its source, or its evidential relations that are relevant to how much credence a claim should receive.

Our primary interest is cue influence: how strongly a cue affects a model's judgment when multiple cues may compete. We measure cue influence behaviorally by holding other relevant features fixed, changing a target cue in a controlled manner, and observing how much that change alters the model's final judgment.
To distinguish weak cue influence from a failure to identify the cue itself, we additionally measure cue availability: whether a model can correctly recover or identify the cue under direct questioning. Here, availability is an operational behavioral measure assessed through explicit identification tasks, rather than a claim about the model's internal representational state. We refer to cases in which cue availability and cue influence diverge as a cue availability--influence dissociation.
This framework also allows us to study multiple potentially competing cues within the same experimental system. Historical source reliability is only one factor that may affect judgment. The quality of current evidence, whether multiple reports derive from independent evidence, the temporal ordering of events, and the presentation order of information may all compete with source history. In particular, multiple speakers expressing the same conclusion do not necessarily constitute multiple independent pieces of evidence: their reports may reflect independent observations, or they may ultimately derive from the same underlying source. Conversely, presentation order may strongly affect model behavior without carrying the same epistemic relevance as source reliability or evidence independence. By manipulating these factors separately, we can ask not only whether a model can identify them, but also which cues actually control its judgment when they compete.
Based on this perspective, we construct a controlled diagnostic benchmark for epistemic cue utilization under uncertainty. 
The benchmark contains 500 evaluation scenarios and includes five complementary diagnostic tracks covering cue representation, integration, reliability utilization, cue competition, and verified reasoning.
Together, these five classes of measurements constitute each model's epistemic cue-utilization profile.
Our cross-model evaluation reveals a clear behavioral dissociation: {being able to identify an epistemic cue does not imply that the cue will influence final judgment to the same degree}. Some models can identify, with near-perfect accuracy, which source has the better reliability history, yet respond very differently when that reliability relationship is reversed. Their final choices range from remaining almost unchanged to nearly completely reversing with the manipulation. This demonstrates that cue identification and cue influence constitute two behavioral dimensions that can vary independently.
This variation is not limited to source reliability. 
These
differences extend beyond source reliability to evidence independence, event
time, and presentation order, yielding distinct cue-utilization profiles
across models. When trusted verification resolves the disputed state, these
differences become much smaller.
This finding raises a natural further question: are these cue-utilization patterns fixed properties of a model, or can they be reshaped by changing how information is presented and used? To address this question, we conduct theoretically motivated interventions that modify the salience and organization of already available information, as well as explicit principles for evidence use, without introducing new evidence. The results show that such interventions can substantially alter the cues on which some models rely, while preserving their factual judgment performance when trusted verification makes the true state explicit.

Our main contributions are:
(1) We introduce a 500-world benchmark that separates whether a model can identify an epistemic cue from whether that cue actually affects its decision.
(2) Across 13 LLMs, we find a clear gap between cue identification and cue use: models that recognize the same evidence can rely on it very differently when making a final judgment.
(3) We show that these cue-use patterns are not fixed. Controlled changes to evidence presentation, organization, and explicit weighting rules can substantially reshape model behavior while preserving strong performance when the true state is directly verified.

\begin{figure*}[t]
\centering
\includegraphics[width=0.95\textwidth]{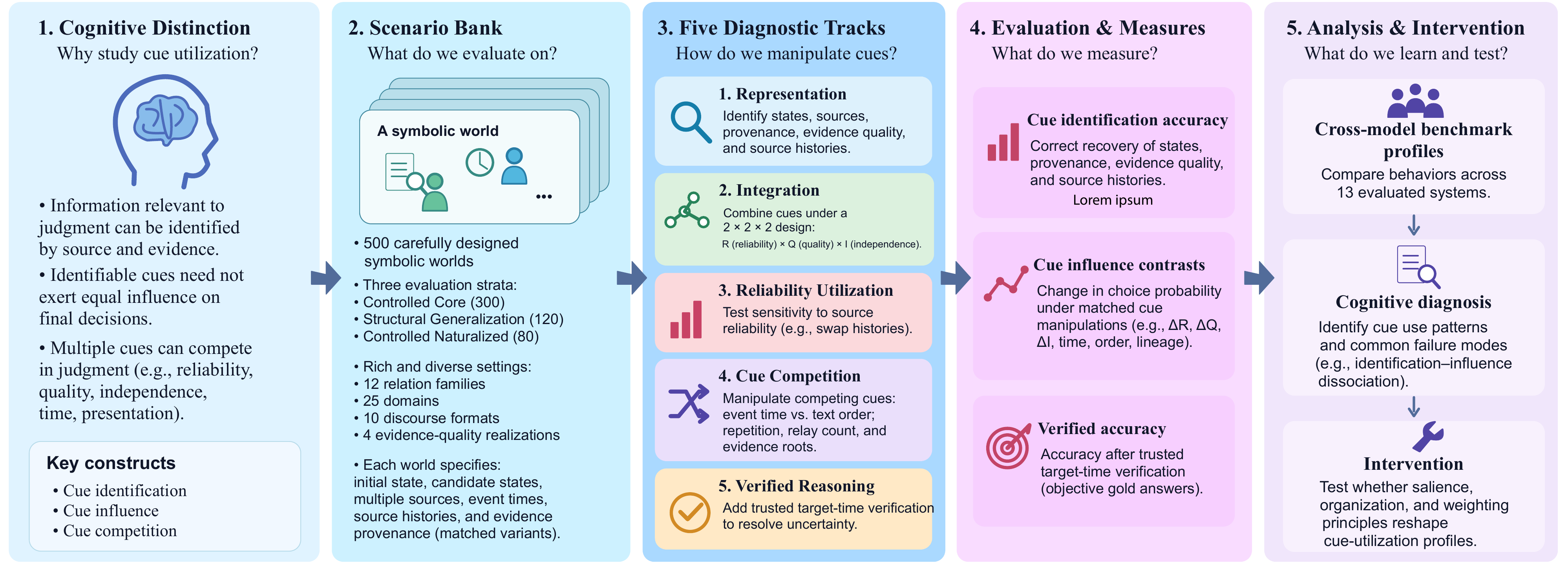}
\vspace{-4mm}
\caption{
Overview of the benchmark. Five diagnostic tracks across 500 evaluation
worlds separate cue identification from cue influence, examine competition
among epistemic cues, and test judgment after trusted verification. The
resulting measurements support cross-model cue-utilization profiles and
controlled intervention analysis.
}
\vspace{-5mm}
\label{zrch}
\end{figure*}
% =========================================================
% 2. Related Work
% =========================================================

% =========================================================
% 3. Benchmark
% =========================================================

\section{Benchmark}
\label{sec:benchmark}

\subsection{Measuring Epistemic Cue Utilization}
\label{sec:benchmark-measurement}

Our benchmark separates two related questions: whether a model can identify a
judgment-relevant cue and how strongly that cue influences its final choice.
For example, a model may correctly identify which source has the better
historical record while barely changing its decision when those histories are
exchanged. We therefore measure {cue identification} and
{cue influence} separately.
Judgment may also depend on several competing cues, including source history,
current evidence quality and independence, event time, and presentation
features. We isolate their influence through matched conditions that vary the
target cue while holding the remaining scenario fixed. This design is
motivated by the cognitive-science distinctions introduced in the
Introduction: source-monitoring research motivates our provenance probes,
epistemic-vigilance work motivates our manipulations of source history and
evidence quality, and cue-utilization research motivates our behavioral
contrasts and cue-competition tests.
When conflicting evidence remains unresolved, support-choice tasks are not
assigned a unique gold state; instead, we measure changes in choice across
matched conditions. The Verified Reasoning track provides the complementary
case in which trusted target-time evidence establishes an objective answer.Figure~\ref{zrch} provides an overview of the benchmark.

\subsection{Scenario Bank and Evaluation Flow}
\label{sec:scenario-bank}

The benchmark contains 500 formal evaluation worlds: 300
\textbf{Controlled Core}, 120 \textbf{Structural Generalization}, and 80
\textbf{Controlled Naturalized} worlds. Evaluation worlds do not overlap with
materials used during benchmark development. Across the 500 worlds, the
benchmark covers 12 relation families, 25 domains, 10 discourse formats,
four evidence-quality realizations, and 10 background-event structures; the
Structural Generalization subset additionally varies five structural axes.
Each world contributes 48 independent requests, yielding 24,000 requests per
system and 312,000 model--request cells across 13 evaluated systems. Full
distributions by relation family, domain, discourse format, and structural
factor are reported in Appendix~\ref{app:dataset-statistics}.

The Controlled Core provides the most tightly matched comparisons. Different
versions of the same world share the target entity, candidate states, event
relations, and source structure, while changing the experimental factor of
interest. Structural Generalization retains the same judgment problems and
experimental variables while varying five structural axes---source count,
evidence-graph topology, dependency pattern, relay-chain length, and role
configuration. The 120 worlds are evenly distributed across these five axes.
Controlled Naturalized worlds preserve the symbolic variables and experimental
controls while expressing them through more naturalized discourse.

Across the full bank, the worlds cover assignment, custody, location,
membership, outcome, permission, priority, quantity, routing, schedule,
state, and version relations. They are distributed across 25 domains and
10 discourse formats. Complete distributions are reported in
Appendix~\ref{app:dataset-statistics}.

Each world is defined by an event structure and an evidence-provenance
structure. We use {source} for an original observer such as $O_1$ or
$O_2$, {evidence record} for an observation produced by that source
($e_1,e_2$), and {relay} for a messenger that faithfully transmits a
record without adding a new observation. {Evidence lineage} denotes the
path from an original evidence record to the final report, while
{source history} refers to how often a source's earlier reports matched
subsequent trusted verification.
Current evidence quality is manipulated through target identifiability rather
than by labeling a report as true or false. In the high-quality condition,
the complete target identifier is available. In the limited-quality
condition, part of the target identifier is unavailable while the reported
state value remains visible. We instantiate this manipulation through textual
descriptions of photographs, logs, documents, and audio records. Exact realizations are provided in
Table~\ref{tab:quality-realizations} in the appendix.
Figure~\ref{example} illustrates a pair of HL/LH conditions from the Reliability Utilization track (\ref{sec:diagnostic-tracks}). All identification probes and final judgments are evaluated in separate model contexts, and the final-choice requests do not include the model's previous responses to any identification probe. Prompt templates and output
protocols are provided in Appendix~\ref{app:protocol1}.

\begin{figure*}[t]
\centering
\includegraphics[width=0.95\textwidth]{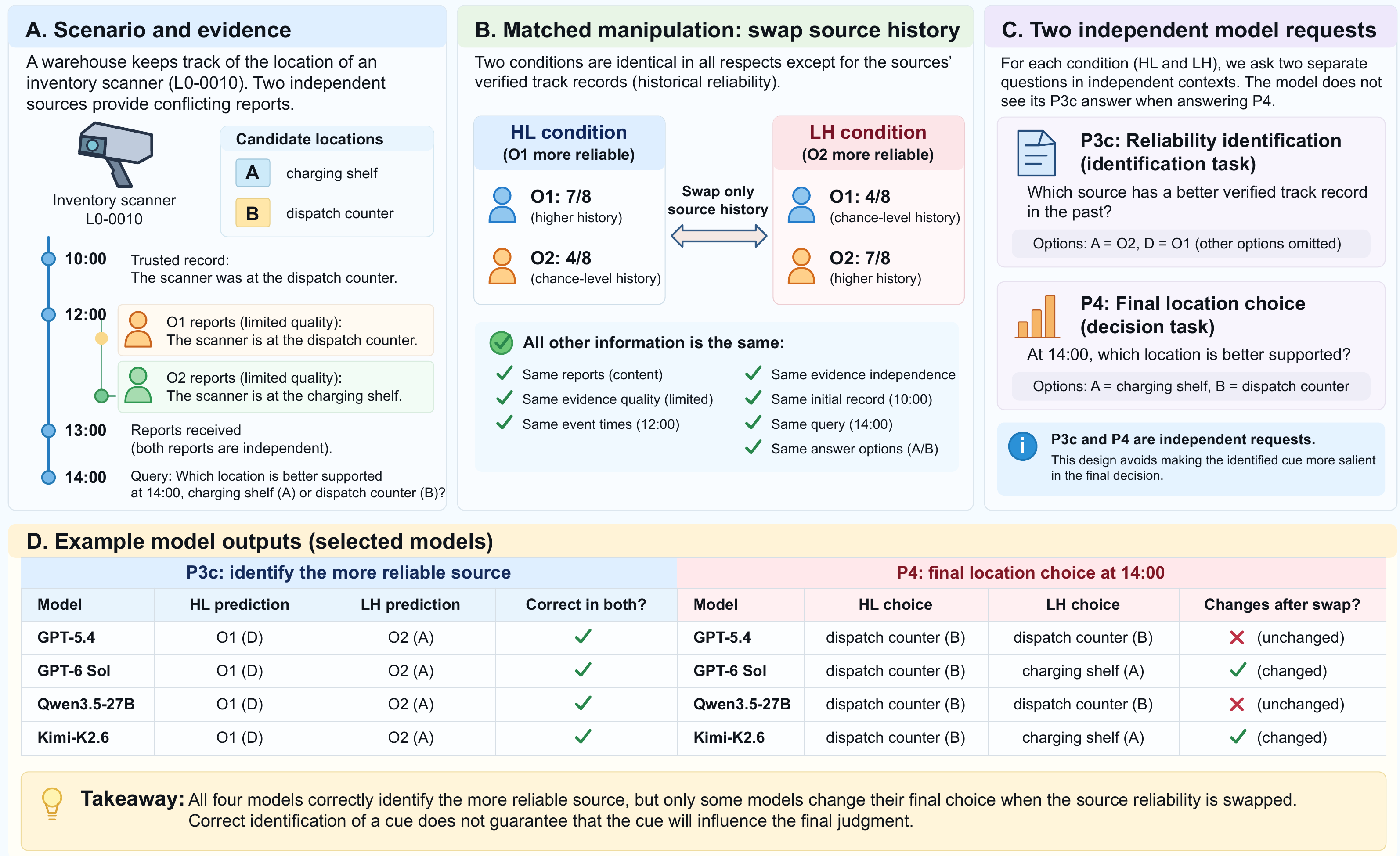}
\vspace{-3mm}
\caption{
Worked example of the reliability-swap diagnostic. Two conflicting sources
support candidate states $A$ and $B$; HL and LH exchange only their source
histories. P3c (relative-reliability identification) and P4 (final support
choice) are issued as separate requests. All four systems shown identify the
historically stronger source, yet their P4 choices respond differently to the
same swap: some are unchanged, others switch to the state backed by the newly
stronger source. P4 has no unique gold answer; the measured quantity is the
change in choice between HL and LH.
}
\vspace{-5mm}
\label{example}
\end{figure*}

\subsection{Diagnostic Tracks}

Table~\ref{tab:track-main}
summarizes details.
\begin{table}[t]
\centering
\tiny
\resizebox{0.5\textwidth}{!}{
\setlength{\tabcolsep}{1pt}
\begin{tabular}{p{2.15cm}p{3.35cm}cp{2.2cm}}
\toprule
\textbf{Track} &
\textbf{Manipulation} &
\textbf{world} &
\textbf{Measure} \\
\midrule

Representation &
State, provenance, source-history probes &
28 &
Identification accuracy \\

Integration &
$R\times Q\times I$ &
8 &
Cue effects, interactions \\

Reliability Utilization &
HL $\leftrightarrow$ LH history swap &
2 &
$\Delta_{\mathrm{hist}}$ \\

Cue Competition &
Time/order and report-lineage controls &
8 &
Cue-specific contrasts \\

Verified Reasoning &
Trusted target-time verification &
2 &
Verified accuracy \\

\bottomrule
\end{tabular}}
\vspace{-2mm}
\caption{
Overview of the five diagnostic tracks.
}
\vspace{-6mm}
\label{tab:track-main}
\end{table}
The benchmark contains five complementary diagnostic tracks
(Table~\ref{tab:track-main}). \textbf{Cue Representation} tests whether
models can recover states, provenance, evidence quality, and source history.
\textbf{Cue Integration} crosses historical reliability ($R$), current
evidence quality ($Q$), and evidence independence ($I$) in a balanced
$2\times2\times2$ design. \textbf{Reliability Utilization} directly tests
whether exchanging the histories of two conflicting sources changes the final
choice. \textbf{Cue Competition} separates event time from text order and
report multiplicity from independent evidence. Finally,
\textbf{Verified Reasoning} adds trusted target-time evidence that resolves
the disputed state. Complete condition mappings and prompt templates are
provided in Appendix~\ref{sec:diagnostic-tracks}.

\subsection{Evaluation Measures}
\label{sec:evaluation-measures}

We report three classes of measures: \textbf{cue identification accuracy},
\textbf{cue influence contrasts}, and \textbf{verified accuracy}. Of the
24,000 requests per system, 15,000 have objective targets and 9,000 are
unverified support-choice requests without a unique gold state. Complete
probe-level request accounting is provided in
Appendix~\ref{app:benchmark-request-composition}.
\textbf{Cue identification accuracy.}
Probes with objective targets are scored by accuracy. Our primary
representation measure is {relative reliability recognition}, which
tests whether the model identifies the better historical source in
unequal-history conditions. Initial-state, provenance and evidence-quality,
and exact-history measures provide additional diagnostics.

\textbf{Cue influence contrasts.}
For unverified support choices, we measure cue influence as the signed
percentage-point change in the empirical proportion of responses selecting
semantic candidate state $B$ between matched conditions. The direction of each
contrast is fixed by the experimental design. Semantic $A/B$ denotes the
underlying candidate states rather than displayed answer positions.
For the reliability swap, we define
\begin{equation}
\small
\Delta_{\mathrm{hist}}
=
\Pr(\hat{Y}=B\mid LH)-\Pr(\hat{Y}=B\mid HL),
\label{eq:history-swap}
\end{equation}
where $\Pr(\hat{Y}=B\mid c)$ denotes the empirical proportion of decisions
selecting state $B$ under condition $c$. This quantity is reported as
{Rel.\ Swap} in Table~\ref{tab:benchmark}. Because O1 supports semantic
state $A$ and O2 supports $B$, a positive value indicates that assigning the
stronger source history to O2 increases selection of $B$.
We compute analogous signed contrasts for evidence quality, factorial
independence, text order, event time, repetition, relay multiplicity, and
evidence lineage. Two- and three-way factorial interactions are computed from
the corresponding contrasts over the matched $R\times Q\times I$ conditions.
\textbf{Verified accuracy.}
In the Verified Reasoning track, a trusted target-time record establishes a
unique target state, and responses are scored by accuracy.
\subsection{Benchmark Results}
\label{sec:benchmark-results}

We evaluate a total of eight open-source models and five proprietary models. We use the Controlled Core as the primary benchmark for model comparison, as it provides the cleanest matched manipulations. More detailed results are provided in Appendix~\ref{app:benchmark-detail}.
\begin{table}[t]
\centering
\resizebox{0.47\textwidth}{!}{
\setlength{\tabcolsep}{3.7pt}
\begin{tabular}{lrrrrrrrr}
\toprule
\textbf{Model} &
\textbf{Valid} &
\textbf{Rel.\ Rec.} &
\textbf{Rel.\ Swap} &
\textbf{$Q$} &
\textbf{$I$} &
\textbf{Text} &
\textbf{Time} &
\textbf{Verified} \\
\midrule
Qwen3.5-2B &
20.8 & 0.0 & -- & -- & -- & -- & -- & 0.0 \\

Qwen3.5-4B &
45.7 & 19.0 & 8.6 & 1.0 & -1.0 & 0.0 & 17.1 & 7.2 \\

Qwen3.5-9B &
100.0 & 87.2 & 0.3 & 13.3 & 13.7 & 20.8 & 32.2 & 100.0 \\

Qwen3.5-27B &
100.0 & 99.2 & 0.7 & 2.5 & 8.7 & -43.7 & 52.3 & 100.0 \\

Qwen3.6-27B &
99.9 & 99.7 & 4.0 & 1.4 & 5.2 & -28.1 & 69.5 & 99.7 \\

Qwen3.8-27B &
100.0 & 99.5 & 0.0 & 0.8 & 0.4 & -8.2 & 44.2 & 100.0 \\

InternVL3.5-8B &
95.9 & 48.8 & 4.7 & 15.1 & 0.9 & 14.0 & 27.3 & 100.0 \\

InternVL3.5-14B &
87.9 & 75.8 & 0.4 & 11.4 & 22.7 & 39.4 & 17.0 & 96.3 \\

\midrule
Grok 4.3 &
99.2 & 100.0 & 73.0 & 12.0 & 18.3 & 5.7 & 58.0 & 98.7 \\

Claude Opus 4.7 &
98.6 & 98.8 & 33.0 & 8.7 & 55.0 & -2.3 & 25.3 & 96.8 \\

GPT-5.4 &
100.0 & 100.0 & 6.3 & 9.2 & 70.2 & -13.8 & 68.2 & 100.0 \\

GPT-6 Sol &
100.0 & 100.0 & 100.0 & 4.8 & 0.7 & -0.2 & 99.8 & 100.0 \\

Kimi-K2.6 &
99.9 & 100.0 & 88.3 & 7.1 & 26.9 & 1.0 & 87.0 & 99.7 \\
\bottomrule
\end{tabular}}
\vspace{-3mm}
\caption{
Main Controlled Core benchmark results across 13 systems.
Rel.\ Rec.\ is the equally weighted mean of relative-reliability
identification accuracy in the HL and LH conditions; Rel.\ Swap denotes
the reliability-history swap effect. $Q$ and $I$ denote evidence quality
and independence, while Text and Time denote presentation order and event
time.
}
\vspace{-6mm}
\label{tab:benchmark}
\end{table}
The clearest cross-model separation appears between relative-reliability recognition and reliability influence. Several systems achieve approximately 99--100\% recognition accuracy on unequal-history probes, yet their responses to the corresponding history swap range from essentially zero to 100 percentage points: Qwen3.8-27B achieves 99.5\% recognition accuracy with a 0.0-point history-swap effect, whereas GPT-6 Sol achieves 100\% recognition accuracy with a 100.0-point swap effect.
The same variation extends beyond source history. Models differ in their sensitivity to current evidence quality, factorial independence, event time, and presentation order. Kimi-K2.6 is highly sensitive to source history (88.3 points) and event time (87.0 points), while showing an approximately zero text-order effect (1.0 point). GPT-5.4 and Claude Opus 4.7 exhibit the largest independence effects (70.2 and 55.0 points, respectively), with small or negative text-order effects. Qwen3.5-27B and InternVL3.5-14B show the largest-magnitude text-order effects ($-43.7$ and 39.4 points, respectively). Figure~\ref{fig:profiles} summarizes these patterns as {cue-utilization profiles}. Detailed results are reported in 
Appendix~\ref{app:benchmark-detail}.
Verified judgments provide a complementary endpoint. Several systems reach
approximately 97--100\% verified accuracy on the Controlled Core while still
showing markedly different reliability, independence, time, and text-order
effects before verification. Kimi-K2.6 reaches 99.7\% verified accuracy while
showing large reliability-swap and event-time effects, further illustrating
that similar verified endpoint performance can coexist with substantially
different patterns of cue sensitivity under uncertainty.
Taken together, the benchmark reveals a central behavioral separation:
information that can be identified reliably need not exert comparable
influence on judgment. At the same time, models with similar verified
performance can rely on very different combinations of source history,
evidence independence, event time, and presentation cues. The next section
examines this identification--influence dissociation more directly through
exact-history reconstruction and targeted cue-competition analyses.

\begin{figure}[t]
\centering
\includegraphics[width=.5\textwidth]{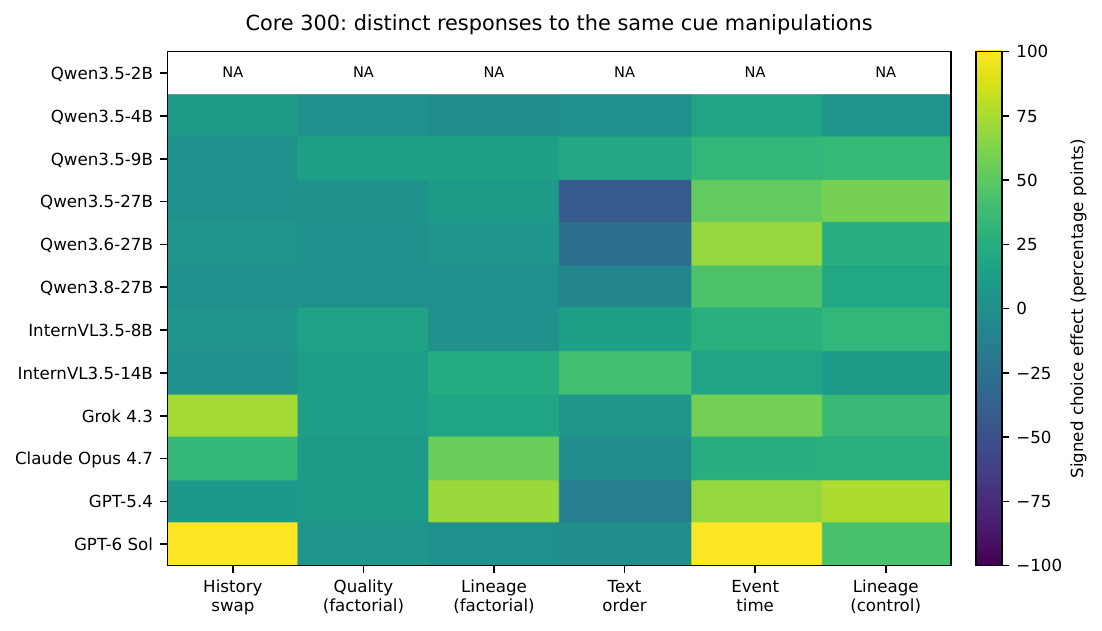}
\vspace{-8mm}
\caption{Core cue-utilization profiles. Columns show distinct experimental contrasts on a common percentage-point display scale, not a shared latent weight. Factorial lineage varies independence within the eight-condition design; the lineage control contrasts two speakers sharing one root with two speakers carrying separate roots. }
\vspace{-5mm}
\label{fig:profiles}
\end{figure}

\section{Experiments}

\label{sec:experiments}

\subsection{Cognitive Analysis and Diagnosis}
\label{sec:diagnosis}

\begin{figure}[t]
\centering
\includegraphics[width=.48\textwidth]{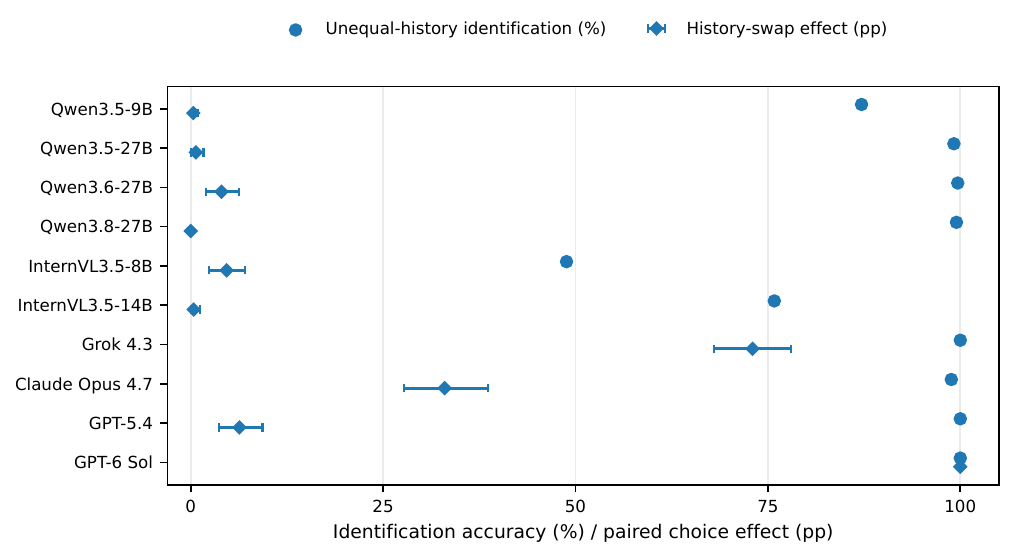}
\vspace{-8mm}
\caption{Source-history identification and influence can dissociate. Circles show mean HL/LH recognition; diamonds show the paired history-swap effect with its supplied 95\% world-bootstrap interval.}
\vspace{-6mm}
\label{fig:recognition-influence}
\end{figure}

\textbf{Recognizing the more reliable source does not determine its influence.}
Source credibility involves both evaluating a communicator and deciding how
much to rely on that communicator's advice \citep{yaniv2000advice}.
Figure~\ref{fig:recognition-influence} separates these two behaviors.
The three 27B Qwen models achieve approximately 99--100\% unequal-history
recognition on Core, yet their history-swap effects range from 0 to only
4.0\pp. Other systems with similarly high recognition show much larger
effects, demonstrating that recognizing the more reliable source does not
determine how strongly source history influences the final choice.
This separation remains when successful identification is conditioned on
directly. For Qwen3.5-27B, restricting the analysis to worlds in which both
independent reliability-identification probes are correct leaves the Core
swap effect essentially unchanged (0.68\pp{} versus 0.67\pp{} overall;
Appendix~\ref{app:identification}). Weak cue influence therefore cannot be
explained simply by failure to identify the cue. This is the central
\emph{identification--influence dissociation}.

\textbf{Source knowledge has more than one resolution.}
The dedicated history diagnostic also separates comparative judgment from exact reconstruction. Qwen3.5-27B and Qwen3.6-27B correctly rank unequal histories on 99.25\% and 99.42\% of trials, whereas recovering {both} exact success counts yields 54.5\% and 58.7\%. Equal histories introduce another distinction: on Core, Qwen3.5-27B's HH and LL recognition is 0\%, despite its near-ceiling unequal-history performance; Grok 4.3 correctly handles all four configurations. These patterns place different demands on source assessment: ranking unequal records, recognizing equality, and reconstructing their exact frequencies are not interchangeable measurements (Tables~\ref{tab:history} and~\ref{tab:counts}).

\textbf{The origin of agreement matters beyond agreement itself.}
Source-monitoring research motivates distinguishing a statement from its origin \citep{johnson1993source}. Our controls illustrate this distinction by varying repetition, speakers, and evidence roots. For Qwen3.5-27B, repeating one report changes choice by 2.0\pp, introducing another speaker sharing that report by 5.7\pp, and replacing the shared root with a separate root by 58.7\pp. GPT-5.4 shows a 75.0\pp{} root effect, while GPT-6 Sol shows 42.3\pp. The latter is much larger than its 0.7\pp{} factorial lineage effect: source dependence has a different consequence in the two evidence configurations. Thus, a profile characterizes responses to specified evidence relations, not a single context-free weight (Appendix~\ref{app:competition}).

\textbf{When an event occurred and where it is written both shape judgment.}
Order-sensitive belief updating is a long-standing problem in judgment research \citep{hogarth1992order}. Crossing event time with presentation order distinguishes temporal relevance from sequence of exposure. Qwen3.5-27B responds by 52.3\pp{} to event time and $-43.7$\pp{} to text order; the corresponding effects are 69.5 and $-28.1$\pp{} for Qwen3.6-27B. GPT-6 Sol instead combines a 99.8\pp{} time effect with a $-0.2$\pp{} order effect. Temporal updating and presentation sensitivity can therefore coexist or dissociate within otherwise competent systems. Appendix~\ref{app:competition} reports the two conditional order effects and the option-mapping controls.
When trusted target-time verification resolves the disputed state, the
higher-performing systems return to near-ceiling accuracy, suggesting that
the preceding differences arise primarily when conflicting evidence still
has to be weighed.

\subsection{Theory-Driven Intervention}
\label{sec:intervention}
The diagnostic distinction raises a further question: is weak influence an enduring property of a cue for a given model, or can the decision context make that same cue consequential? We test this by intervening on the presentation and use of existing evidence, rather than adding a new observation.

\begin{figure}[t]
\centering
\includegraphics[width=.23\textwidth]{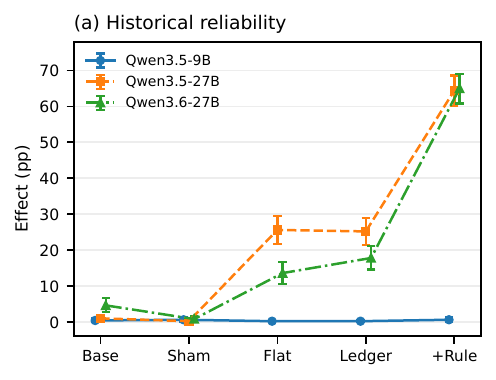}\hfill
\includegraphics[width=.23\textwidth]{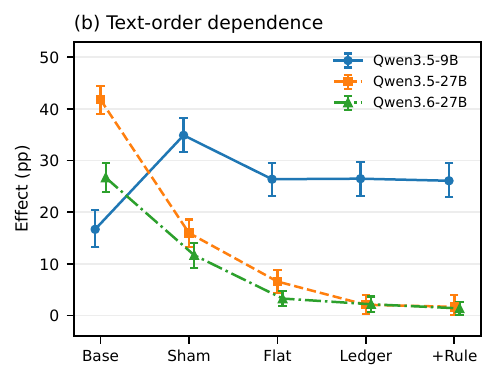}\\[-1mm]
\includegraphics[width=.23\textwidth]{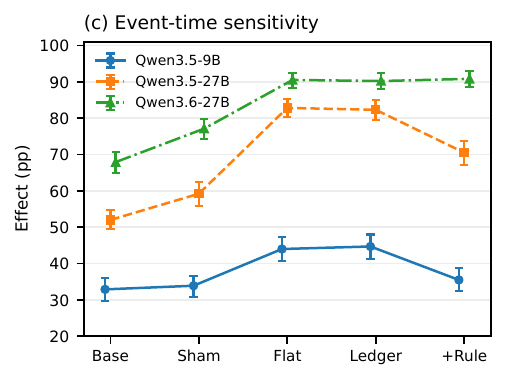}\hfill
\includegraphics[width=.23\textwidth]{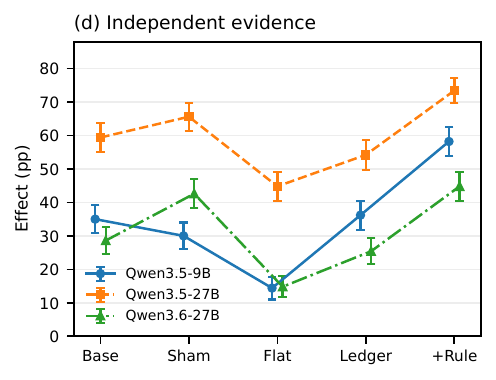}
\vspace{-3mm}
\caption{Interventions reshape different parts of the cue profile. whiskers are the supplied 95\% world-bootstrap intervals. Text-order dependence is $|\mathbb{E}_w[\Delta_{\mathrm{text},w}]|$.}
\vspace{-6mm}
\label{fig:intervention}
\end{figure}

\textbf{Restatement, organization, and a rule for use.}
We evaluate all 500 worlds with Qwen3.5-9B, Qwen3.5-27B, and
Qwen3.6-27B under five intervention conditions. \textbf{Base} (C0)
uses the original input, while \textbf{Sham} (C1) adds length-matched
but evidentially irrelevant structure. \textbf{Flat} (C2) explicitly
restates the relevant source, history, quality, lineage, and event-time
facts in prose. \textbf{Ledger} (C3) presents the same fact set as C2,
but organizes it by source and evidence relation. \textbf{+Rule} (C4)
adds a fixed, answer-independent evidence-weighting principle to the
ledger. This design separates the effects of making relevant facts
explicit, organizing those facts, and providing a principle for how
they should influence judgment.
Each condition contains 20 decision requests per world, yielding
150,000 requests across the three models; Base is rerun within the
same intervention experiment. We pre-specify four component contrasts:
C2--C0 tests explicit restatement relative to the original input,
C3--C2 isolates the effect of organization given the same facts,
C3--C1 compares the ledger against a length-matched control, and
C4--C3 tests the additional weighting rule. The primary outcomes are
the reliability-swap effect and the absolute text-order effect.
Across three models, four contrasts, two outcomes, and the Core-300
and All-500 analysis sets, this gives a family of
$3\times4\times2\times2=48$ pre-specified primary tests, corrected
jointly using Holm's procedure. Figure~\ref{fig:intervention}
summarizes the resulting cue-utilization profiles, and
Table~\ref{tab:intervention-contrasts} reports the All-500 component
contrasts.

\textbf{Restating a cue can make it consequential.}
Flat raises the history-swap effect from 1.0 to 25.6\pp{} in Qwen3.5-27B and from 4.6 to 13.6\pp{} in Qwen3.6-27B. Both changes are significant in the pre-specified comparison family. These are changes in the effect of {swapping histories}, not simply a greater tendency to choose one state. The contrast links the diagnosis to an experimental intervention: a source history that has little influence in the original narrative can become consequential when its evidential information is made explicit again at the point of judgment.

\textbf{Organization and a weighting principle make different contributions.}
Ledger alone does not explain this reliability response. Relative to Flat, its change is $-0.4$\pp{} for Qwen3.5-27B and $+4.2$\pp{} for Qwen3.6-27B; neither passes the corrected main comparison. Adding the rule to Ledger, however, raises history sensitivity by 39.2 and 47.2\pp, reaching 64.4 and 65.0\pp. The decisive reliability contrast is therefore not prose versus a table-like record, but the further specification of how existing evidence should enter a judgment. Organization still changes other dimensions: Ledger increases the separate-root contrast over Flat by 9.4 and 10.6\pp{} in the two 27B models (Table~\ref{tab:intervention-secondary}).

\begin{table}[t]
\centering
\resizebox{0.47\textwidth}{!}{
\fontsize{8}{9.3}\selectfont
\setlength{\tabcolsep}{2.5pt}

\begin{tabular}{lrrr}
\toprule
\textbf{Model} &
\textbf{Contrast} &
$\Delta$ Reliability &
$\Delta |\mathrm{TextOrder}|$ \\
\midrule

Qwen3.5
& C3$-$C2 & $+0.0\;[-0.6, 0.6]$ & $+0.1\;[-2.5, 2.6]$ \\
\textit{9B}
& C3$-$C1 & $-0.4\;[-1.2, 0.4]$ & $-8.4\;[-11.4, -5.4]^{*}$ \\
&
C4$-$C3 & $+0.4\;[0.0, 1.0]$ & $-0.4\;[-3.2, 2.4]$ \\
&
C2$-$C0 & $-0.2\;[-0.6, 0.0]$ & $+9.7\;[6.4, 12.8]^{*}$ \\

\cmidrule(lr){1-4}

Qwen3.5
& C3$-$C2 & $-0.4\;[-3.0, 2.2]$ & $-4.5\;[-6.5, -2.5]^{*}$ \\
\textit{27B}
& C3$-$C1 & $+25.0\;[21.2, 28.8]^{*}$ & $-13.9\;[-16.7, -11.1]^{*}$ \\
&
C4$-$C3 & $+39.2\;[34.8, 43.4]^{*}$ & $-0.4\;[-2.5, 1.8]$ \\
&
C2$-$C0 & $+24.6\;[21.0, 28.4]^{*}$ & $-35.2\;[-38.1, -32.2]^{*}$ \\

\cmidrule(lr){1-4}

Qwen3.6
& C3$-$C2 & $+4.2\;[1.2, 7.2]$ & $-1.1\;[-2.5, 0.4]$ \\
\textit{27B}
& C3$-$C1 & $+17.0\;[13.8, 20.4]^{*}$ & $-9.5\;[-11.8, -7.1]^{*}$ \\
&
C4$-$C3 & $+47.2\;[42.8, 51.6]^{*}$ & $-0.8\;[-2.0, 0.4]$ \\
&
C2$-$C0 & $+9.0\;[6.0, 12.0]^{*}$ & $-23.4\;[-26.2, -20.6]^{*}$ \\

\bottomrule
\end{tabular}}
\vspace{-3mm}
\caption{
Pre-specified intervention contrasts on the  500-world set.
Entries are percentage-point differences in the reliability-swap effect
and absolute text-order effect, with 95\% world-bootstrap confidence intervals.
$^{*}$ denotes Holm-corrected significance at $p<.05$.
}
\vspace{-6mm}
\label{tab:intervention-contrasts}
\end{table}

\textbf{Order sensitivity responds to more than evidential structure.}
Sham already reduces absolute order effects from 41.8 to 16.0\pp{} and from 26.7 to 11.7\pp{} in the two 27B models. Flat reduces them further to 6.6 and 3.3\pp. Ledger adds a corrected reduction of 4.5\pp{} over Flat in Qwen3.5-27B, but not in Qwen3.6-27B. This separates a broad context effect from the incremental contribution of organizing evidence. The conditional early- and late-event results follow the same reduction pattern for the 27B restatement contrast (Table~\ref{tab:intervention-order}); it is not an artifact of averaging opposite order effects.

\textbf{Plasticity is specific to both the model and the cue.}
Qwen3.5-9B maintains a history-swap effect of only 0.2--0.6\pp{} across all treatments. It is nevertheless responsive to the intervention: +Rule increases its lineage effect from 36.2 to 58.2\pp{} relative to Ledger. Its absolute text-order effect rises above Base under all four appended-input conditions. The contrast with 27B is therefore not simply that one model is affected and the other is not; different cues are affected.
The 27B shifts are multidimensional. In Qwen3.5-27B, +Rule strengthens both history and lineage responses but reduces the time effect by 11.8\pp{} relative to Ledger, while remaining above Base (70.5 versus 52.0\pp). Qwen3.6-27B retains a 90.8\pp{} time effect alongside its  reliability increase. Verified accuracy remains 100\% for every model and treatment. Together, these results show that identifying a cue, making it influential, and changing its influence are empirically distinct questions. The intervention changes the pattern of evidence use, rather than uniformly improving every entry in that pattern.

\section{Related Work}

\textbf{Conflicting information and belief updating in LLMs.}
Prior work has examined conflicts among contextual information, parametric knowledge, and external information sources \citep{xu2024knowledge,ying2024intuitive,du2024context,tan2024blinded}.
Other studies have investigated whether updated or conflicting information propagates through subsequent reasoning and how to increase model reliance on contextual evidence \citep{feng2026tracking,zhou2023contextfaithful,wang2025adacad}.
Related work also shows that model judgments can be influenced by social pressure and task-irrelevant context \citep{zhu2025conformity,gonen2025semantic}.
Building on these studies, we separately measure whether a model can identify a judgment-relevant cue and how much a matched manipulation of that cue changes its final decision.
\textbf{Sources, credibility, and epistemic judgment.}
Source-monitoring research distinguishes informational content from its attributed origin \citep{johnson1993source}, while studies of source credibility and advice taking examine how source properties and prior performance influence reliance on a source \citep{hovland1951influence,birnbaum1976combining,yaniv2000advice}.
Epistemic-vigilance accounts similarly emphasize evaluating both communicated content and source trustworthiness \citep{sperber2010epistemic}.
Recent studies have also begun to examine the effects of source reliability and source labels on model judgments \citep{cheng2026accommodation,sun2026label}.
Our study further examines how source history, evidence provenance, and evidence independence influence model judgments.
\textbf{Cue integration and presentation effects.}
Research on multiple-cue judgment examines how decision makers integrate several partially informative cues and assign them different functional weights \citep{slovic1971comparison,brehmer1972cue,anderson1981foundations}.
Judgments can also be influenced by the order and position in which information is presented \citep{hogarth1992order,liu2024lost,zhou2026curse}.
Based on these findings, we distinguish event time from textual presentation order and repeated reports from independent evidence to characterize how models utilize competing cues.

\vspace{-1mm}
\section{Conclusion}
\vspace{-1mm}
We introduce a benchmark separating cue identification from cue influence. Across models, these behaviors often diverge, producing distinct patterns of cue use. Controlled interventions show that these patterns are malleable: changing how evidence is presented and used can reshape which cues influence decisions while preserving strong factual performance when the true state is explicit.
This suggests that evaluating language models under uncertainty requires more than measuring final answers or answer changes. It requires distinguishing what models can identify, what influences their judgments, and whether that influence can be changed.

\section{Limitations} 

Our benchmark is intentionally highly controlled. The 500 evaluation worlds are constructed textual scenarios, allowing source history, evidence quality, evidence lineage, event time, and presentation order to be manipulated independently. This level of control helps isolate the effects of individual cues, but it does not fully capture the complexity of real-world evidence environments, including naturally occurring information, long interactive contexts, and multimodal inputs. Although the Controlled Naturalized subset varies the surface form of the scenarios, these materials remain researcher-controlled rather than independently authored real-world cases. Future work should examine whether the same cue-utilization patterns also emerge in naturally occurring and multimodal settings.
The intervention study is also narrower in scope than the benchmark evaluation. It is conducted only on three Qwen checkpoints and uses prompt-level interventions involving fact restatement, information organization, and evidence-weighting rules. The observed reweighting therefore shows that cue use can be changed in these models, but we have not yet established whether the same effects generalize to other model architectures.

\section{Ethical Considerations}

Our benchmark uses fictional, text-based scenarios rather than real
individuals' private records. It measures how language models respond
to conflicting evidence, but cue sensitivity should not be interpreted
as a measure of factual correctness or a guarantee of trustworthy
judgment. In particular, stronger reliance on a particular cue does
not necessarily imply better decisions in real-world settings.
Our findings may help researchers identify weaknesses in how models
assess source reliability, evidence independence, and conflicting
information. However, knowledge of these sensitivities could also
be misused to steer model judgments through selective evidence
presentation or misleading source descriptions. We therefore view
the benchmark as a tool for evaluating and improving model behavior,
rather than as a substitute for independent evidence verification
in consequential decisions.
‘
\bibliography{custom}

\begin{thebibliography}{22}
\providecommand{\natexlab}[1]{#1}

\bibitem[{Anderson(1981)}]{anderson1981foundations}
Norman~H. Anderson. 1981.
\newblock \emph{Foundations of Information Integration Theory}.
\newblock Academic Press, New York.

\bibitem[{Birnbaum et~al.(1976)Birnbaum, Wong, and Wong}]{birnbaum1976combining}
Michael~H. Birnbaum, Rebecca Wong, and Leighton~K. Wong. 1976.
\newblock \href {https://doi.org/10.3758/BF03213185} {Combining information from sources that vary in credibility}.
\newblock \emph{Memory \& Cognition}, 4(3):330--336.

\bibitem[{Brehmer(1972)}]{brehmer1972cue}
Berndt Brehmer. 1972.
\newblock \href {https://doi.org/10.1016/0030-5073(72)90051-7} {Cue utilization and cue consistency in multiple-cue probability learning}.
\newblock \emph{Organizational Behavior and Human Performance}, 8(2):286--296.

\bibitem[{Cheng et~al.(2026)Cheng, Hawkins, and Jurafsky}]{cheng2026accommodation}
Myra Cheng, Robert~D Hawkins, and Dan Jurafsky. 2026.
\newblock Accommodation and epistemic vigilance: A pragmatic account of why llms fail to challenge harmful beliefs.
\newblock In \emph{Proceedings of the 64th Annual Meeting of the Association for Computational Linguistics (Volume 1: Long Papers)}, pages 16181--16203.

\bibitem[{Du et~al.(2024)Du, Sn{\ae}bjarnarson, Stoehr, White, Schein, and Cotterell}]{du2024context}
Kevin Du, V{\'e}steinn Sn{\ae}bjarnarson, Niklas Stoehr, Jennifer White, Aaron Schein, and Ryan Cotterell. 2024.
\newblock Context versus prior knowledge in language models.
\newblock In \emph{Proceedings of the 62nd Annual Meeting of the Association for Computational Linguistics (Volume 1: Long Papers)}, pages 13211--13235.

\bibitem[{Feng et~al.(2026)Feng, Chen, Wu, Zhou, and Bosselut}]{feng2026tracking}
Yiyang Feng, Zeming Chen, Haotian Wu, Jiawei Zhou, and Antoine Bosselut. 2026.
\newblock Tracking the limits of knowledge propagation: How llms fail at multi-step reasoning with conflicting knowledge.
\newblock In \emph{Proceedings of the 19th Conference of the European Chapter of the Association for Computational Linguistics (Volume 1: Long Papers)}, pages 5813--5847.

\bibitem[{Gonen et~al.(2025)Gonen, Blevins, Liu, Zettlemoyer, and Smith}]{gonen2025semantic}
Hila Gonen, Terra Blevins, Alisa Liu, Luke Zettlemoyer, and Noah~A Smith. 2025.
\newblock Does liking yellow imply driving a school bus? semantic leakage in language models.
\newblock In \emph{Proceedings of the 2025 Conference of the Nations of the Americas Chapter of the Association for Computational Linguistics: Human Language Technologies (Volume 1: Long Papers)}, pages 785--798.

\bibitem[{Hogarth and Einhorn(1992)}]{hogarth1992order}
Robin~M. Hogarth and Hillel~J. Einhorn. 1992.
\newblock \href {https://doi.org/10.1016/0010-0285(92)90002-J} {Order effects in belief updating: The belief-adjustment model}.
\newblock \emph{Cognitive Psychology}, 24(1):1--55.

\bibitem[{Hovland and Weiss(1951)}]{hovland1951influence}
Carl~I. Hovland and Walter Weiss. 1951.
\newblock \href {https://doi.org/10.1086/266350} {The influence of source credibility on communication effectiveness}.
\newblock \emph{Public Opinion Quarterly}, 15(4):635--650.

\bibitem[{Johnson et~al.(1993)Johnson, Hashtroudi, and Lindsay}]{johnson1993source}
Marcia~K. Johnson, Shahin Hashtroudi, and D.~Stephen Lindsay. 1993.
\newblock \href {https://doi.org/10.1037/0033-2909.114.1.3} {Source monitoring}.
\newblock \emph{Psychological Bulletin}, 114(1):3--28.

\bibitem[{Liu et~al.(2024)Liu, Lin, Hewitt, Paranjape, Bevilacqua, Petroni, and Liang}]{liu2024lost}
Nelson~F Liu, Kevin Lin, John Hewitt, Ashwin Paranjape, Michele Bevilacqua, Fabio Petroni, and Percy Liang. 2024.
\newblock Lost in the middle: How language models use long contexts.
\newblock \emph{Transactions of the association for computational linguistics}, 12:157--173.

\bibitem[{Slovic and Lichtenstein(1971)}]{slovic1971comparison}
Paul Slovic and Sarah Lichtenstein. 1971.
\newblock Comparison of bayesian and regression approaches to the study of information processing in judgment.
\newblock \emph{Organizational behavior and human performance}, 6(6):649--744.

\bibitem[{Sperber et~al.(2010)Sperber, Cl{\'e}ment, Heintz, Mascaro, Mercier, Origgi, and Wilson}]{sperber2010epistemic}
Dan Sperber, Fabrice Cl{\'e}ment, Christophe Heintz, Olivier Mascaro, Hugo Mercier, Gloria Origgi, and Deirdre Wilson. 2010.
\newblock \href {https://doi.org/10.1111/j.1468-0017.2010.01394.x} {Epistemic vigilance}.
\newblock \emph{Mind \& Language}, 25(4):359--393.

\bibitem[{Sun et~al.(2026)Sun, Wu, Qin, Echizen, El~Ali, and Sugawara}]{sun2026label}
Xin Sun, Di~Wu, Sijing Qin, Isao Echizen, Abdallah El~Ali, and Saku Sugawara. 2026.
\newblock Label effects: Shared heuristic reliance in trust assessment by humans and llm-as-a-judge.
\newblock In \emph{Proceedings of the 64th Annual Meeting of the Association for Computational Linguistics (Volume 1: Long Papers)}, pages 32378--32392.

\bibitem[{Tan et~al.(2024)Tan, Sun, Yang, Wang, Cao, and Cheng}]{tan2024blinded}
Hexiang Tan, Fei Sun, Wanli Yang, Yuanzhuo Wang, Qi~Cao, and Xueqi Cheng. 2024.
\newblock Blinded by generated contexts: How language models merge generated and retrieved contexts when knowledge conflicts?
\newblock In \emph{Proceedings of the 62nd Annual Meeting of the Association for Computational Linguistics (Volume 1: Long Papers)}, pages 6207--6227.

\bibitem[{Wang et~al.(2025)Wang, Prasad, Stengel-Eskin, and Bansal}]{wang2025adacad}
Han Wang, Archiki Prasad, Elias Stengel-Eskin, and Mohit Bansal. 2025.
\newblock Adacad: Adaptively decoding to balance conflicts between contextual and parametric knowledge.
\newblock In \emph{Proceedings of the 2025 Conference of the Nations of the Americas Chapter of the Association for Computational Linguistics: Human Language Technologies (Volume 1: Long Papers)}, pages 11636--11652.

\bibitem[{Xu et~al.(2024)Xu, Qi, Guo, Wang, Wang, Zhang, and Xu}]{xu2024knowledge}
Rongwu Xu, Zehan Qi, Zhijiang Guo, Cunxiang Wang, Hongru Wang, Yue Zhang, and Wei Xu. 2024.
\newblock Knowledge conflicts for llms: A survey.
\newblock In \emph{Proceedings of the 2024 Conference on Empirical Methods in Natural Language Processing}, pages 8541--8565.

\bibitem[{Yaniv and Kleinberger(2000)}]{yaniv2000advice}
Ilan Yaniv and Eli Kleinberger. 2000.
\newblock \href {https://doi.org/10.1006/obhd.2000.2909} {Advice taking in decision making: Egocentric discounting and reputation formation}.
\newblock \emph{Organizational Behavior and Human Decision Processes}, 83(2):260--281.

\bibitem[{Ying et~al.(2024)Ying, Cao, Xiong, Cui, He, and Liu}]{ying2024intuitive}
Jiahao Ying, Yixin Cao, Kai Xiong, Long Cui, Yidong He, and Yongbin Liu. 2024.
\newblock Intuitive or dependent? investigating llms’ behavior style to conflicting prompts.
\newblock In \emph{Proceedings of the 62nd Annual Meeting of the Association for Computational Linguistics (Volume 1: Long Papers)}, pages 4221--4246.

\bibitem[{Zhou et~al.(2023)Zhou, Zhang, Poon, and Chen}]{zhou2023contextfaithful}
Wenxuan Zhou, Sheng Zhang, Hoifung Poon, and Muhao Chen. 2023.
\newblock Context-faithful prompting for large language models.
\newblock In \emph{Findings of the Association for Computational Linguistics: EMNLP 2023}, pages 14544--14556.

\bibitem[{Zhou et~al.(2026)Zhou, Zou, Di~Eugenio, and Zhang}]{zhou2026curse}
Yue Zhou, Henry~Peng Zou, Barbara Di~Eugenio, and Yang Zhang. 2026.
\newblock The curse of verbalization: How presentation order constrains llm reasoning.
\newblock In \emph{Findings of the Association for Computational Linguistics: EACL 2026}, pages 4175--4185.

\bibitem[{Zhu et~al.(2025)Zhu, Zhang, Stafford, Collier, and Vlachos}]{zhu2025conformity}
Xiaochen Zhu, Caiqi Zhang, Tom Stafford, Nigel Collier, and Andreas Vlachos. 2025.
\newblock Conformity in large language models.
\newblock In \emph{Proceedings of the 63rd Annual Meeting of the Association for Computational Linguistics (Volume 1: Long Papers)}, pages 3854--3872.

\end{thebibliography}
\clearpage
\appendix

\section{Experiments details}

\subsection{Diagnostic Tracks setting}
\label{sec:diagnostic-tracks}

\textbf{Cue Representation.}
The first track examines whether a model can recover the information required for subsequent judgment. The probes cover the trusted initial state, the relationship between messages and their underlying evidence records, current evidence quality, and source history. The reliability probes distinguish exact history reconstruction from relative comparison: P3a and P3b require the model to report the specific numbers of verified matches for $O_1$ and $O_2$, respectively, whereas P3c asks which source had the higher proportion of correct reports in its prior history.

\textbf{Cue Integration.}
The second track uses a balanced $2\times2\times2$ design over historical
reliability ($R$), current evidence quality ($Q$), and evidence independence
($I$). For $R$, sources have either a higher-history profile, with 7 of 8
earlier reports matching trusted verification, or a chance-level-history
profile, with 4 of 8 matching verification. $Q$ varies whether the target
identifier in the current evidence is complete or only partially available.
$I$ varies whether the current reports trace to distinct original evidence
records or to a shared original record. The full eight-condition mapping is
reported in Appendix~\ref{app:factorial-prompts}.

\textbf{Reliability Utilization.}
The third track provides a direct behavioral counterpart to relative
reliability identification. Two sources support opposite semantic states
while their current reports, evidence quality, timing, and evidence roots are
held fixed. In HL, the source supporting semantic state $A$ has a 7/8 history
and the source supporting $B$ has a 4/8 history; LH reverses these histories.
The corresponding P3c probe measures which source is identified as
historically better, whereas the final-choice pair measures how much the same
swap changes judgment.

\textbf{Cue Competition.}
The fourth track contains two sets of controls. The first orthogonalizes event time and textual presentation order, separating sensitivity to {when the event occurred} from sensitivity to {where the report appears in the prompt}. The second is a report-structure sequence that progressively disentangles the number of messages, the number of relays, and the number of underlying evidence records. The key lineage contrast holds constant two messages and two relays, varying only whether they share a single underlying evidence record or originate from two independent records. The complete report-structure conditions are provided in Appendix~\ref{app:report-structure}.

\textbf{Verified Reasoning.}
The final track adds a separate trusted record that directly establishes the
target state at 14:00. Verified-A and Verified-B differ only in the value
established by this final record. Because the target-time state is explicit,
these requests have objective gold answers and are evaluated by accuracy.

\subsection{Protocols, Denominators, and Source Alignment}
\label{app:protocol}
The result sets have three distinct request universes. The unified benchmark evaluates 13 LLMs with 24,000 requests each. The dedicated six-Qwen diagnostic study evaluates 41,020 requests per model, using selected scenario subsets for some follow-up tasks. The intervention evaluates three models with 50,000 requests each. Shared scenario names do not make these request sets identical. All main cross-system profiles come from the unified benchmark; diagnostic and intervention tables identify their own scopes.

A benchmark scenario generates 28 representation requests (eight initial-state, eight provenance, four relative-history, and eight single-source count requests), eight factorial choices, two history-swap choices, eight order/report-structure choices, and two verified-state choices. These sum to 48. The six Qwen benchmark runs reuse 19,120 observations per model where inputs and inference identity match, and add 4,880 new requests per model. Identification and verified accuracies use the planned denominator, while choice contrasts require valid paired responses.

\paragraph{Model scope.}
The local group contains six Qwen checkpoints and two InternVL3.5 checkpoints, evaluated with text input. The InternVL models use a Qwen3 language backbone. The four API labels are the model identifiers recorded in the benchmark. Local systems use the frozen greedy/non-thinking profile and API systems use provider-native settings. Eight local and four API systems are scored.

\paragraph{Intervals.}
The unified benchmark reports 5,000 whole-world bootstrap intervals. The dedicated diagnostic study retains its frozen 2,000-resample analyses. The intervention uses at least 5,000 resamples and a fixed 48-comparison Holm family. Figures reproduce supplied intervals. The HL/LH point average used for recognition has equal planned branch sizes; no pooled recognition interval is synthesized from the two marginal intervals.

\paragraph{Direction.}
History swap is LH minus HL; Q is clear minus limited; I is separate minus shared roots. Text order is B-first minus A-first, averaged across event times; event time is B-at-12:00 minus B-at-08:00, averaged across presentation orders. Repetition, speaker multiplicity, and lineage are successive paired contrasts. These probability-scale effects retain their individual task definitions.
\begin{table}[t]
\centering
\resizebox{0.5\textwidth}{!}{
\setlength{\tabcolsep}{4pt}
\begin{tabular}{lrrrrr}
\toprule
System & Planned & Valid & Valid (\%) & Tech. fail & Trunc. flag \\
\midrule
Qwen3.5-2B & 24000 & 5002 & 20.8 & 0 & 0 \\
Qwen3.5-4B & 24000 & 10972 & 45.7 & 0 & 0 \\
Qwen3.5-9B & 24000 & 24000 & 100.0 & 0 & 0 \\
Qwen3.5-27B & 24000 & 24000 & 100.0 & 0 & 0 \\
Qwen3.6-27B & 24000 & 23972 & 99.9 & 28 & 0 \\
Qwen3.8-27B & 24000 & 24000 & 100.0 & 0 & 0 \\
InternVL3.5-8B & 24000 & 23005 & 95.9 & 0 & 0 \\
InternVL3.5-14B & 24000 & 21094 & 87.9 & 0 & 555 \\
Grok 4.3 & 24000 & 23814 & 99.2 & 0 & 0 \\
Claude Opus 4.7 & 24000 & 23658 & 98.6 & 0 & 0 \\
GPT-5.4 & 24000 & 24000 & 100.0 & 0 & 0 \\
GPT-6 Sol & 24000 & 24000 & 100.0 & 0 & 0 \\
\bottomrule
\end{tabular}}
\caption{Unified benchmark completion and response validity. Non-valid response flags retain the source accounting; they are not added again to the planned denominator.}
\label{tab:coverage}
\end{table}

\subsection{Full Benchmark Profiles}
\label{app:benchmark-detail}
The tables below preserve recognition, choice effects, and output coverage as separate measurements. The main profile focuses on Core; the other splits expose variation across the materials rather than supply a replacement primary analysis.
\begin{table}[t]
\centering
\resizebox{0.5\textwidth}{!}{
\setlength{\tabcolsep}{4pt}
\begin{tabular}{lrrrrrrrr}
\toprule
Model & Valid & Rec. & Sw. & Q & I & Text & Time & Ver. \\
\midrule
Qwen3.5-2B$^{\dagger}$ & 20.8 & 0.0 & \textemdash & \textemdash & \textemdash & \textemdash & \textemdash & 0.0 \\
Qwen3.5-4B$^{\dagger}$ & 45.7 & 24.7 & 7.0 & 1.1 & -2.2 & 0.3 & 14.5 & 8.5 \\
Qwen3.5-9B & 100.0 & 85.8 & 0.6 & 11.2 & 14.8 & 16.6 & 32.4 & 100.0 \\
Qwen3.5-27B & 100.0 & 98.7 & 1.2 & 2.8 & 9.3 & -41.8 & 51.8 & 100.0 \\
Qwen3.6-27B & 99.9 & 99.3 & 4.2 & 2.3 & 5.6 & -26.4 & 68.2 & 99.8 \\
Qwen3.8-27B & 100.0 & 99.2 & 0.0 & 0.8 & -0.4 & -7.7 & 42.7 & 100.0 \\
InternVL3.5-8B & 95.9 & 49.8 & 7.0 & 13.9 & 0.4 & 13.4 & 26.4 & 100.0 \\
InternVL3.5-14B$^{\dagger}$ & 87.9 & 73.9 & 0.5 & 11.7 & 23.0 & 38.1 & 17.6 & 93.5 \\
\midrule
Grok 4.3 & 99.2 & 100.0 & 74.2 & 10.6 & 19.7 & 6.5 & 58.7 & 98.3 \\
Claude Opus 4.7 & 98.6 & 99.0 & 32.1 & 9.4 & 50.8 & -1.4 & 25.3 & 96.8 \\
GPT-5.4 & 100.0 & 100.0 & 7.8 & 9.0 & 69.8 & -12.7 & 68.9 & 100.0 \\
GPT-6 Sol & 100.0 & 99.9 & 100.0 & 5.1 & 1.4 & 0.1 & 99.7 & 100.0 \\
\bottomrule
\end{tabular}}
\caption{Benchmark profiles on all 500 scenarios. Valid, Rec., and Ver. are percentages; all other columns are signed percentage-point effects. Rec. is the equally weighted mean of HL/LH unequal-history judgments; Sw. is the reliability-history swap, Q and I are factorial effects, and Text and Time are the orthogonal order/time contrasts. Ver. averages the two verified states. Each subset retains its own valid-pair denominator.}
\label{tab:all-profile}
\end{table}
\begin{table}[t]
\centering
\resizebox{0.5\textwidth}{!}{
\setlength{\tabcolsep}{4pt}
\begin{tabular}{lrrrrrrrr}
\toprule
Model & Valid & Rec. & Sw. & Q & I & Text & Time & Ver. \\
\midrule
Qwen3.5-2B$^{\dagger}$ & 21.7 & 0.0 & \textemdash & \textemdash & \textemdash & \textemdash & \textemdash & 0.0 \\
Qwen3.5-4B$^{\dagger}$ & 40.0 & 24.2 & 5.1 & 0.0 & -5.0 & 0.0 & 12.2 & 5.0 \\
Qwen3.5-9B & 100.0 & 87.1 & 0.8 & 10.8 & 17.9 & 12.5 & 32.5 & 100.0 \\
Qwen3.5-27B & 100.0 & 98.3 & 0.8 & 3.1 & 14.8 & -43.8 & 55.4 & 100.0 \\
Qwen3.6-27B & 99.9 & 99.6 & 5.0 & 2.5 & 5.9 & -30.4 & 67.1 & 100.0 \\
Qwen3.8-27B & 100.0 & 100.0 & 0.0 & 1.2 & -1.7 & -10.8 & 45.8 & 100.0 \\
InternVL3.5-8B & 96.9 & 54.2 & 5.8 & 11.2 & 4.6 & 11.7 & 29.2 & 100.0 \\
InternVL3.5-14B$^{\dagger}$ & 88.5 & 75.0 & 0.0 & 13.7 & 22.3 & 40.5 & 14.5 & 90.4 \\
\midrule
Grok 4.3 & 99.3 & 100.0 & 74.2 & 9.2 & 19.3 & 10.0 & 60.8 & 96.7 \\
Claude Opus 4.7 & 98.6 & 99.6 & 32.5 & 7.7 & 52.9 & -0.8 & 21.7 & 95.4 \\
GPT-5.4 & 100.0 & 100.0 & 5.8 & 7.9 & 68.3 & -12.1 & 65.4 & 100.0 \\
GPT-6 Sol & 100.0 & 100.0 & 100.0 & 4.2 & 1.7 & 0.4 & 99.6 & 100.0 \\
\bottomrule
\end{tabular}}
\caption{Benchmark profiles on Structural 120. Valid, Rec., and Ver. are percentages; all other columns are signed percentage-point effects. Rec. is the equally weighted mean of HL/LH unequal-history judgments; Sw. is the reliability-history swap, Q and I are factorial effects, and Text and Time are the orthogonal order/time contrasts. Ver. averages the two verified states. Each subset retains its own valid-pair denominator.}
\label{tab:structural-profile}
\end{table}
\begin{table}[t]
\centering
\resizebox{0.5\textwidth}{!}{
\setlength{\tabcolsep}{4pt}
\begin{tabular}{lrrrrrrrr}
\toprule
Model & Valid & Rec. & Sw. & Q & I & Text & Time & Ver. \\
\midrule
Qwen3.5-2B$^{\dagger}$ & 21.1 & 0.0 & \textemdash & \textemdash & \textemdash & \textemdash & \textemdash & 0.0 \\
Qwen3.5-4B$^{\dagger}$ & 61.2 & 46.9 & 4.3 & 2.2 & -2.2 & 1.0 & 11.0 & 18.8 \\
Qwen3.5-9B & 100.0 & 78.8 & 1.2 & 4.1 & 14.1 & 6.9 & 33.1 & 100.0 \\
Qwen3.5-27B & 100.0 & 97.5 & 3.8 & 3.1 & 3.8 & -31.9 & 44.4 & 100.0 \\
Qwen3.6-27B & 100.0 & 97.5 & 3.8 & 5.3 & 6.6 & -13.8 & 65.0 & 100.0 \\
Qwen3.8-27B & 100.0 & 96.9 & 0.0 & 0.0 & -1.2 & -1.2 & 32.5 & 100.0 \\
InternVL3.5-8B & 96.9 & 46.9 & 17.5 & 13.1 & -8.1 & 13.8 & 18.8 & 100.0 \\
InternVL3.5-14B$^{\dagger}$ & 85.1 & 65.0 & 1.9 & 9.9 & 25.7 & 28.8 & 25.4 & 87.5 \\
\midrule
Grok 4.3 & 99.5 & 100.0 & 78.8 & 7.5 & 25.6 & 4.4 & 58.1 & 99.4 \\
Claude Opus 4.7 & 98.9 & 98.8 & 27.8 & 14.6 & 32.1 & 1.3 & 30.8 & 98.8 \\
GPT-5.4 & 100.0 & 100.0 & 16.2 & 10.0 & 70.6 & -9.4 & 76.9 & 100.0 \\
GPT-6 Sol & 100.0 & 99.4 & 100.0 & 7.2 & 3.4 & 0.6 & 99.4 & 100.0 \\
\bottomrule
\end{tabular}}
\caption{Benchmark profiles on Naturalistic 80. Valid, Rec., and Ver. are percentages; all other columns are signed percentage-point effects. Rec. is the equally weighted mean of HL/LH unequal-history judgments; Sw. is the reliability-history swap, Q and I are factorial effects, and Text and Time are the orthogonal order/time contrasts. Ver. averages the two verified states. Each subset retains its own valid-pair denominator.}
\label{tab:natural-profile}
\end{table}
\begin{table*}[t]
\centering
\footnotesize
\setlength{\tabcolsep}{4pt}
\begin{tabular}{lrrrr}
\toprule
Model & History swap & R & Q & I \\
\midrule
Qwen3.5-2B & $\text{--}$ & $\text{--}$ & $\text{--}$ & $\text{--}$ \\
Qwen3.5-4B & $8.6\;[4.3, 13.7]$ & $-3.0\;[-7.0, -0.5]$ & $1.0\;[-2.0, 5.0]$ & $-1.0\;[-3.0, 1.0]$ \\
Qwen3.5-9B & $0.3\;[0.0, 1.0]$ & $-6.7\;[-9.5, -4.0]$ & $13.3\;[10.8, 16.1]$ & $13.7\;[9.7, 17.4]$ \\
Qwen3.5-27B & $0.7\;[0.0, 1.7]$ & $-0.8\;[-1.8, 0.0]$ & $2.5\;[1.4, 3.8]$ & $8.7\;[6.0, 11.7]$ \\
Qwen3.6-27B & $4.0\;[2.0, 6.3]$ & $-0.9\;[-1.8, -0.1]$ & $1.4\;[0.5, 2.5]$ & $5.2\;[3.2, 7.3]$ \\
Qwen3.8-27B & $0.0\;[0.0, 0.0]$ & $-0.2\;[-0.8, 0.2]$ & $0.8\;[0.2, 1.5]$ & $0.4\;[0.0, 1.0]$ \\
InternVL3.5-8B & $4.7\;[2.3, 7.0]$ & $-6.8\;[-9.3, -4.2]$ & $15.1\;[12.2, 17.9]$ & $0.9\;[-2.0, 3.8]$ \\
InternVL3.5-14B & $0.4\;[0.0, 1.2]$ & $-9.7\;[-13.2, -6.5]$ & $11.4\;[8.2, 14.6]$ & $22.7\;[17.7, 27.7]$ \\
Grok 4.3 & $73.0\;[68.0, 78.0]$ & $48.7\;[44.8, 52.7]$ & $12.0\;[9.1, 14.9]$ & $18.3\;[15.3, 21.5]$ \\
Claude Opus 4.7 & $33.0\;[27.7, 38.7]$ & $23.7\;[21.3, 26.1]$ & $8.7\;[6.4, 10.8]$ & $55.0\;[51.6, 58.3]$ \\
GPT-5.4 & $6.3\;[3.7, 9.3]$ & $10.9\;[8.6, 13.2]$ & $9.2\;[6.8, 11.8]$ & $70.2\;[67.0, 73.5]$ \\
GPT-6 Sol & $100.0\;[100.0, 100.0]$ & $94.3\;[92.9, 95.7]$ & $4.8\;[3.6, 6.2]$ & $0.7\;[-0.6, 1.9]$ \\
\bottomrule
\end{tabular}
\caption{Core history and factorial main effects (pp; 95\% intervals).}
\label{tab:core-effects}
\end{table*}
\begin{table*}[t]
\centering
\footnotesize
\setlength{\tabcolsep}{4pt}
\begin{tabular}{lrrr}
\toprule
Model & Text order & Event time & Separate-root control \\
\midrule
Qwen3.5-2B & $\text{--}$ & $\text{--}$ & $\text{--}$ \\
Qwen3.5-4B & $0.0\;[-5.2, 4.8]$ & $17.1\;[11.4, 23.3]$ & $4.6\;[0.8, 9.2]$ \\
Qwen3.5-9B & $20.8\;[15.8, 25.7]$ & $32.2\;[28.3, 36.2]$ & $34.0\;[28.7, 39.7]$ \\
Qwen3.5-27B & $-43.7\;[-47.2, -40.3]$ & $52.3\;[49.0, 55.7]$ & $58.7\;[53.0, 64.3]$ \\
Qwen3.6-27B & $-28.1\;[-31.6, -24.7]$ & $69.5\;[66.0, 72.9]$ & $26.0\;[21.3, 31.3]$ \\
Qwen3.8-27B & $-8.2\;[-10.7, -5.7]$ & $44.2\;[39.3, 49.2]$ & $20.0\;[15.7, 24.7]$ \\
InternVL3.5-8B & $14.0\;[9.8, 18.0]$ & $27.3\;[23.7, 31.2]$ & $32.7\;[27.3, 38.0]$ \\
InternVL3.5-14B & $39.4\;[34.6, 44.6]$ & $17.0\;[14.0, 20.0]$ & $8.8\;[5.5, 12.5]$ \\
Grok 4.3 & $5.7\;[3.2, 8.3]$ & $58.0\;[53.3, 62.7]$ & $35.0\;[29.7, 40.7]$ \\
Claude Opus 4.7 & $-2.3\;[-5.5, 0.8]$ & $25.3\;[21.5, 29.3]$ & $27.0\;[22.3, 32.0]$ \\
GPT-5.4 & $-13.8\;[-16.7, -11.0]$ & $68.2\;[64.0, 72.3]$ & $75.0\;[70.0, 80.0]$ \\
GPT-6 Sol & $-0.2\;[-0.5, 0.0]$ & $99.8\;[99.5, 100.0]$ & $42.3\;[36.7, 48.0]$ \\
\bottomrule
\end{tabular}
\caption{Core temporal, presentation, and lineage-control effects (pp; 95\% intervals).}
\label{tab:core-context}
\end{table*}

\subsection{Source Identification and Historical Comparisons}
\label{app:identification}
The main recognition measure concerns unequal histories. Table~\ref{tab:history} retains the two equal-history controls and the combined four-condition score. Table~\ref{tab:representation} additionally separates initial-state identification, provenance, and individual count extraction. These scores answer different questions from the joint exact-count diagnostic in Table~\ref{tab:counts}.
\begin{table}[t]
\centering
\resizebox{0.5\textwidth}{!}{
\setlength{\tabcolsep}{4pt}
\begin{tabular}{lrrrrr}
\toprule
Model & HH & LL & HL & LH & All four \\
\midrule
Qwen3.5-2B & 0.0 & 0.0 & 0.0 & 0.0 & 0.0 \\
Qwen3.5-4B & 0.0 & 0.0 & 20.0 & 18.0 & 9.5 \\
Qwen3.5-9B & 0.3 & 0.0 & 92.0 & 82.3 & 43.7 \\
Qwen3.5-27B & 0.0 & 0.0 & 99.7 & 98.7 & 49.6 \\
Qwen3.6-27B & 0.0 & 49.3 & 100.0 & 99.3 & 62.2 \\
Qwen3.8-27B & 0.3 & 3.0 & 100.0 & 99.0 & 50.6 \\
InternVL3.5-8B & 3.7 & 0.0 & 79.3 & 18.3 & 25.3 \\
InternVL3.5-14B & 0.0 & 0.0 & 94.3 & 57.3 & 37.9 \\
Grok 4.3 & 100.0 & 100.0 & 100.0 & 100.0 & 100.0 \\
Claude Opus 4.7 & 31.7 & 49.7 & 100.0 & 97.7 & 69.8 \\
GPT-5.4 & 4.3 & 83.0 & 100.0 & 100.0 & 71.8 \\
GPT-6 Sol & 99.0 & 99.0 & 100.0 & 100.0 & 99.5 \\
\bottomrule
\end{tabular}}
\caption{Historical-source identification on Core (\%). H denotes 7/8 verified reports and L denotes 4/8. HH and LL require recognizing equal historical performance; HL and LH require ordering the two sources.}
\label{tab:history}
\end{table}
\begin{table*}[t]
\centering
\footnotesize
\setlength{\tabcolsep}{4pt}
\begin{tabular}{lrrrr}
\toprule
Model & Initial (\%) & Provenance (\%) & Count (\%) & Swap after correct P3c (pp) \\
\midrule
Qwen3.5-2B & 0.0 & 14.8 & 21.2 & $\text{--}$ \\
Qwen3.5-4B & 7.4 & 19.2 & 49.4 & $0.0\;[0.0, 0.0]$ \\
Qwen3.5-9B & 100.0 & 89.6 & 46.0 & $0.4\;[0.0, 1.3]$ \\
Qwen3.5-27B & 100.0 & 100.0 & 70.4 & $0.7\;[0.0, 1.7]$ \\
Qwen3.6-27B & 99.8 & 99.8 & 75.8 & $4.0\;[2.0, 6.4]$ \\
Qwen3.8-27B & 100.0 & 100.0 & 60.2 & $0.0\;[0.0, 0.0]$ \\
InternVL3.5-8B & 94.2 & 64.3 & 16.5 & $13.8\;[3.4, 27.6]$ \\
InternVL3.5-14B & 100.0 & 78.9 & 28.9 & $0.6\;[0.0, 1.9]$ \\
Grok 4.3 & 100.0 & 99.2 & 94.6 & $73.0\;[68.0, 78.0]$ \\
Claude Opus 4.7 & 97.4 & 97.4 & 87.1 & $33.1\;[28.0, 38.6]$ \\
GPT-5.4 & 100.0 & 100.0 & 94.8 & $6.3\;[3.7, 9.3]$ \\
GPT-6 Sol & 100.0 & 100.0 & 99.6 & $100.0\;[100.0, 100.0]$ \\
\bottomrule
\end{tabular}
\caption{Additional Core representation scores and the descriptive history-swap contrast among pairs whose two independent P3c judgments are correct. Count here is the benchmark individual count score, not joint reconstruction of both sources.}
\label{tab:representation}
\end{table*}
\begin{table*}[t]
\centering
\footnotesize
\setlength{\tabcolsep}{4pt}
\begin{tabular}{llrr}
\toprule
Model & P3c correct/planned & Unequal-history P3c (\%) & Joint exact count (\%) \\
\midrule
Qwen3.5-2B & 0/1200 & $0.0\;[0.0, 0.0]$ & $6.3\;[5.2, 7.4]$ \\
Qwen3.5-4B & 266/1200 & $22.2\;[19.0, 25.3]$ & $23.5\;[22.6, 24.5]$ \\
Qwen3.5-9B & 1027/1200 & $85.6\;[83.2, 87.8]$ & $21.3\;[20.1, 22.5]$ \\
Qwen3.5-27B & 1191/1200 & $99.2\;[98.5, 99.8]$ & $54.5\;[52.3, 56.7]$ \\
Qwen3.6-27B & 1193/1200 & $99.4\;[98.8, 99.9]$ & $58.7\;[56.4, 60.9]$ \\
Qwen3.8-27B & 1194/1200 & $99.5\;[98.8, 100.0]$ & $36.2\;[34.1, 38.3]$ \\
\bottomrule
\end{tabular}
\caption{Dedicated six-Qwen history diagnostic on 300 selected worlds. The P3c denominator is 1,200 unequal-history requests; joint exact reconstruction requires two correct count answers in a pair, averaged within world.}
\label{tab:counts}
\end{table*}

\subsection{Cue Competition and the Verified Boundary}
\label{app:competition}
Table~\ref{tab:competition} separates repetition, the number of speakers, and the number of evidence roots. Table~\ref{tab:order-simple} retains order effects at each event time, so a small averaged effect can be distinguished from cancellation. The complete condition-level source export also retains the answer-mapping controls. These controls and response coverage are particularly important for the smaller models.
\begin{table*}[t]
\centering
\footnotesize
\setlength{\tabcolsep}{4pt}
\begin{tabular}{lrrr}
\toprule
Model & Repetition & Another speaker & Separate root \\
\midrule
Qwen3.5-2B & $\text{--}$ & $\text{--}$ & $\text{--}$ \\
Qwen3.5-4B & $-0.6\;[-3.5, 2.3]$ & $1.4\;[-2.1, 4.8]$ & $4.6\;[0.8, 9.2]$ \\
Qwen3.5-9B & $-2.7\;[-6.7, 1.0]$ & $7.3\;[4.7, 10.3]$ & $34.0\;[28.7, 39.7]$ \\
Qwen3.5-27B & $2.0\;[-2.3, 6.3]$ & $5.7\;[2.7, 8.7]$ & $58.7\;[53.0, 64.3]$ \\
Qwen3.6-27B & $4.3\;[-0.7, 9.4]$ & $14.0\;[10.0, 18.1]$ & $26.0\;[21.3, 31.3]$ \\
Qwen3.8-27B & $-7.0\;[-10.7, -3.7]$ & $8.3\;[5.3, 11.7]$ & $20.0\;[15.7, 24.7]$ \\
InternVL3.5-8B & $-1.3\;[-4.3, 1.3]$ & $3.0\;[0.7, 5.3]$ & $32.7\;[27.3, 38.0]$ \\
InternVL3.5-14B & $0.0\;[-1.2, 1.2]$ & $-0.4\;[-1.1, 0.0]$ & $8.8\;[5.5, 12.5]$ \\
Grok 4.3 & $-2.7\;[-6.7, 1.3]$ & $6.7\;[2.3, 11.0]$ & $35.0\;[29.7, 40.7]$ \\
Claude Opus 4.7 & $17.3\;[12.0, 22.7]$ & $9.7\;[5.0, 14.3]$ & $27.0\;[22.3, 32.0]$ \\
GPT-5.4 & $2.7\;[0.3, 5.3]$ & $1.3\;[-1.3, 4.0]$ & $75.0\;[70.0, 80.0]$ \\
GPT-6 Sol & $-23.0\;[-28.3, -17.7]$ & $-1.3\;[-6.7, 4.0]$ & $42.3\;[36.7, 48.0]$ \\
\bottomrule
\end{tabular}
\caption{Core report-structure contrasts (pp; 95\% intervals). Each column changes a different property of a report or its provenance.}
\label{tab:competition}
\end{table*}
\begin{table*}[t]
\centering
\footnotesize
\setlength{\tabcolsep}{4pt}
\begin{tabular}{lrrr}
\toprule
Model & Order: early event & Order: late event & Order $	imes$ time \\
\midrule
Qwen3.5-2B & $\text{--}$ & $\text{--}$ & $\text{--}$ \\
Qwen3.5-4B & $-2.3\;[-7.8, 3.1]$ & $0.7\;[-5.1, 6.6]$ & $5.7\;[-2.9, 14.3]$ \\
Qwen3.5-9B & $31.0\;[24.7, 37.0]$ & $10.7\;[5.0, 16.3]$ & $-20.3\;[-27.3, -13.7]$ \\
Qwen3.5-27B & $-49.3\;[-55.0, -43.7]$ & $-38.0\;[-43.3, -32.7]$ & $11.3\;[2.3, 20.0]$ \\
Qwen3.6-27B & $-29.5\;[-34.9, -24.5]$ & $-26.4\;[-31.4, -21.4]$ & $3.0\;[-4.7, 10.4]$ \\
Qwen3.8-27B & $-12.7\;[-17.7, -8.0]$ & $-3.7\;[-6.0, -1.7]$ & $9.0\;[3.7, 14.3]$ \\
InternVL3.5-8B & $25.0\;[20.0, 30.3]$ & $3.0\;[-2.0, 7.7]$ & $-22.0\;[-28.0, -16.0]$ \\
InternVL3.5-14B & $30.5\;[25.1, 36.3]$ & $48.8\;[42.6, 55.0]$ & $17.2\;[10.4, 23.6]$ \\
Grok 4.3 & $0.7\;[0.0, 1.7]$ & $10.7\;[5.7, 16.0]$ & $10.0\;[5.0, 15.3]$ \\
Claude Opus 4.7 & $-1.3\;[-7.3, 4.7]$ & $-3.3\;[-6.0, -0.7]$ & $-2.0\;[-8.7, 4.7]$ \\
GPT-5.4 & $-1.7\;[-3.3, 0.0]$ & $-26.0\;[-31.7, -20.7]$ & $-24.3\;[-30.3, -18.7]$ \\
GPT-6 Sol & $0.0\;[0.0, 0.0]$ & $-0.3\;[-1.0, 0.0]$ & $-0.3\;[-1.0, 0.0]$ \\
\bottomrule
\end{tabular}
\caption{Conditional Core text-order contrasts and their interaction with event time (pp; 95\% intervals).}
\label{tab:order-simple}
\end{table*}
\begin{table}[t]
\centering
\resizebox{0.5\textwidth}{!}{
\setlength{\tabcolsep}{4pt}
\begin{tabular}{lrrr}
\toprule
Model & Semantic errors & Valid $K$ cases & Contributing worlds \\
\midrule
Qwen3.5-2B & \textemdash & 0 & 0 \\
Qwen3.5-4B & \textemdash & 0 & 0 \\
Qwen3.5-9B & 0 & 39 & 22 \\
Qwen3.5-27B & 0 & 380 & 198 \\
Qwen3.6-27B & 0 & 380 & 197 \\
Qwen3.8-27B & 0 & 303 & 155 \\
\bottomrule
\end{tabular}}
\caption{Dedicated verified diagnostic. K requires three independent checks to be correct. Denominator zero denotes no qualifying cases, not perfect accuracy. The 3.6-27B run has one further K-qualified case with an invalid/missing decision.}
\label{tab:verified-k}
\end{table}

\subsection{Factorial Interactions and Predictive Checks}
\label{app:factorial}
The pilot's large quality--independence interaction motivated the frozen six-Qwen Core test. It was not confirmed at scale: all five estimable Holm-adjusted tests have $p=1.0$, and 2B is not estimable. The cross-system benchmark retains the full set of interaction estimates below. In the dedicated study, adding interaction terms also changes held-out log loss by less than 0.0013 in every estimable Core comparison (Table~\ref{tab:cv}). This distinguishes the robust differences in cue profiles from a universal quality--independence interaction.
\begin{table*}[t]
\centering
\footnotesize
\setlength{\tabcolsep}{4pt}
\begin{tabular}{lrrrr}
\toprule
Model & $R\times Q$ & $R\times I$ & $Q\times I$ & $R\times Q\times I$ \\
\midrule
Qwen3.5-2B & $\text{--}$ & $\text{--}$ & $\text{--}$ & $\text{--}$ \\
Qwen3.5-4B & $2.0\;[-2.0, 6.0]$ & $-2.0\;[-6.0, 2.0]$ & $2.0\;[-2.0, 6.0]$ & $-4.0\;[-12.0, 4.0]$ \\
Qwen3.5-9B & $-1.7\;[-5.0, 1.5]$ & $-3.7\;[-8.0, 0.7]$ & $3.0\;[-1.7, 7.7]$ & $-4.0\;[-10.7, 2.3]$ \\
Qwen3.5-27B & $0.3\;[-1.0, 1.7]$ & $1.3\;[-0.2, 2.8]$ & $-1.3\;[-3.7, 1.2]$ & $-2.7\;[-5.3, 0.0]$ \\
Qwen3.6-27B & $0.5\;[-0.9, 1.9]$ & $1.9\;[0.5, 3.4]$ & $-0.2\;[-2.4, 2.0]$ & $-2.4\;[-5.4, 0.7]$ \\
Qwen3.8-27B & $0.5\;[-0.3, 1.7]$ & $-0.2\;[-0.5, 0.0]$ & $0.5\;[-0.3, 1.7]$ & $0.3\;[0.0, 1.0]$ \\
InternVL3.5-8B & $2.2\;[-1.3, 5.5]$ & $3.2\;[0.2, 6.0]$ & $-1.2\;[-4.7, 2.7]$ & $-1.0\;[-6.7, 5.0]$ \\
InternVL3.5-14B & $2.2\;[-1.4, 5.7]$ & $-1.1\;[-5.9, 3.5]$ & $4.3\;[-1.4, 10.0]$ & $2.2\;[-5.4, 9.7]$ \\
Grok 4.3 & $4.7\;[-0.2, 9.5]$ & $26.7\;[20.3, 32.7]$ & $-0.7\;[-5.7, 4.2]$ & $-9.3\;[-18.3, 0.0]$ \\
Claude Opus 4.7 & $2.4\;[-1.4, 6.2]$ & $-37.7\;[-43.3, -32.2]$ & $-15.2\;[-19.6, -10.9]$ & $-2.1\;[-10.0, 5.9]$ \\
GPT-5.4 & $-3.5\;[-7.7, 0.5]$ & $7.2\;[2.0, 12.2]$ & $-2.2\;[-7.2, 2.7]$ & $-12.3\;[-21.0, -4.0]$ \\
GPT-6 Sol & $0.0\;[-2.5, 2.7]$ & $9.7\;[7.2, 12.2]$ & $-8.7\;[-11.2, -6.3]$ & $0.7\;[-4.0, 5.3]$ \\
\bottomrule
\end{tabular}
\caption{Complete Core interaction estimates in the unified benchmark (pp; 95\% intervals). These estimates retain their original probability-scale contrast definitions.}
\label{tab:interactions}
\end{table*}
\begin{table}[t]
\centering
\resizebox{0.5\textwidth}{!}{
\setlength{\tabcolsep}{4pt}
\begin{tabular}{lrrrr}
\toprule
Model & Main log loss & +Int. log loss & Main Brier & +Int. Brier \\
\midrule
Qwen3.5-2B & \textemdash & \textemdash & \textemdash & \textemdash \\
Qwen3.5-4B & 0.529715 & 0.530920 & 0.172794 & 0.173110 \\
Qwen3.5-9B & 0.664369 & 0.664052 & 0.236040 & 0.235918 \\
Qwen3.5-27B & 0.246970 & 0.247009 & 0.065520 & 0.065533 \\
Qwen3.6-27B & 0.213164 & 0.213214 & 0.052840 & 0.052841 \\
Qwen3.8-27B & 0.089265 & 0.089219 & 0.016889 & 0.016888 \\
\bottomrule
\end{tabular}}
\caption{Frozen five-fold Core world-grouped predictive comparison in the dedicated six-Qwen diagnostic study. Lower is better; both models use the same folds and regularization.}
\label{tab:cv}
\end{table}

\subsection{Intervention: Complete Component Comparisons}
\label{app:intervention}
Each intervention includes all 500 scenarios, and all 150,000 formal requests are valid. Base, Sham, Flat, Ledger, and +Rule use the same underlying scenario evidence. Flat and Ledger are generated from an identical fact set; +Rule retains the ledger verbatim. Sham differs from Ledger by no more than 10\% in token length under the three model tokenizers. Both verified states receive the same appendix, excluding answer-specific summaries.

The primary 48-comparison family consists of three models, two scopes (All/Core), four contrasts, and two endpoints (signed history sensitivity and the absolute aggregate order effect). Tables~\ref{tab:intervention-contrasts} and~\ref{tab:intervention-core-contrasts} cover that family. Component effects on time, repetition, speakers, and lineage are reported with their supplied pointwise intervals in the following tables.
\begin{table*}[t]
\centering
\footnotesize
\setlength{\tabcolsep}{4pt}
\begin{tabular}{llrrrrrrrr}
\toprule
Model & Cond. & Hist. & Q & I & Text & Time & Rep. & Spk. & Root \\
\midrule
Qwen3.5-9B & C0 & 0.4 & 11.2 & 15.2 & 16.7 & 32.9 & -3.0 & 8.0 & 35.0 \\
Qwen3.5-9B & C1 & 0.6 & 10.8 & 20.8 & 34.9 & 33.9 & -6.0 & 11.8 & 30.0 \\
Qwen3.5-9B & C2 & 0.2 & 7.8 & 14.8 & 26.4 & 44.0 & 8.0 & 12.2 & 14.4 \\
Qwen3.5-9B & C3 & 0.2 & 19.4 & 23.9 & 26.5 & 44.7 & -8.0 & 11.8 & 36.2 \\
Qwen3.5-9B & C4 & 0.6 & 18.9 & 30.9 & 26.1 & 35.5 & -8.4 & 8.2 & 58.2 \\
\midrule
Qwen3.5-27B & C0 & 1.0 & 2.8 & 9.3 & -41.8 & 52.0 & 1.2 & 6.6 & 59.4 \\
Qwen3.5-27B & C1 & 0.2 & 4.0 & 9.2 & -16.0 & 59.2 & 4.0 & 5.4 & 65.6 \\
Qwen3.5-27B & C2 & 25.6 & 5.4 & 6.7 & -6.6 & 82.8 & -26.2 & 10.6 & 44.8 \\
Qwen3.5-27B & C3 & 25.2 & 10.7 & 15.7 & -2.1 & 82.3 & -12.2 & 12.6 & 54.2 \\
Qwen3.5-27B & C4 & 64.4 & 16.0 & 25.1 & -1.7 & 70.5 & -17.2 & 13.6 & 73.4 \\
\midrule
Qwen3.6-27B & C0 & 4.6 & 2.0 & 6.1 & -26.7 & 67.9 & 3.2 & 15.6 & 28.6 \\
Qwen3.6-27B & C1 & 0.8 & 2.9 & 7.3 & -11.7 & 77.1 & 7.2 & 21.0 & 42.6 \\
Qwen3.6-27B & C2 & 13.6 & 2.9 & 3.9 & -3.3 & 90.5 & -2.6 & 10.4 & 14.8 \\
Qwen3.6-27B & C3 & 17.8 & 10.4 & 15.4 & -2.2 & 90.2 & -9.2 & 16.6 & 25.4 \\
Qwen3.6-27B & C4 & 65.0 & 14.0 & 24.0 & -1.4 & 90.8 & -1.6 & 17.4 & 44.8 \\
\bottomrule
\end{tabular}
\caption{Complete selected cue profiles on All 500 (signed pp). Verified accuracy and valid-response rate are 100\% in every cell. C0--C4 are Base, Sham, Flat, Ledger, and +Rule.}
\label{tab:intervention-profile}
\end{table*}
\begin{table*}[t]
\centering
\footnotesize
\setlength{\tabcolsep}{4pt}
\begin{tabular}{llrrrrrrrr}
\toprule
Model & Cond. & Hist. & Q & I & Text & Time & Rep. & Spk. & Root \\
\midrule
Qwen3.5-9B & C0 & 0.0 & 13.2 & 14.2 & 21.0 & 32.7 & -2.3 & 6.7 & 34.3 \\
Qwen3.5-9B & C1 & 0.3 & 11.8 & 23.2 & 43.2 & 35.8 & -6.7 & 9.3 & 31.7 \\
Qwen3.5-9B & C2 & 0.0 & 8.8 & 15.2 & 33.0 & 47.0 & 10.0 & 10.0 & 15.3 \\
Qwen3.5-9B & C3 & 0.0 & 21.9 & 27.4 & 32.5 & 44.2 & -9.0 & 12.0 & 34.7 \\
Qwen3.5-9B & C4 & 0.3 & 20.8 & 33.6 & 31.8 & 38.5 & -9.0 & 7.7 & 60.3 \\
\midrule
Qwen3.5-27B & C0 & 0.3 & 2.6 & 8.4 & -44.3 & 52.0 & 2.0 & 5.7 & 58.7 \\
Qwen3.5-27B & C1 & 0.0 & 4.2 & 9.7 & -16.3 & 61.7 & 3.7 & 4.0 & 67.3 \\
Qwen3.5-27B & C2 & 23.7 & 6.1 & 6.2 & -8.0 & 84.7 & -29.0 & 10.3 & 44.7 \\
Qwen3.5-27B & C3 & 23.3 & 10.6 & 16.9 & -3.2 & 84.5 & -14.7 & 12.0 & 54.7 \\
Qwen3.5-27B & C4 & 65.0 & 15.5 & 24.8 & -3.3 & 70.7 & -20.7 & 15.0 & 73.0 \\
\midrule
Qwen3.6-27B & C0 & 4.3 & 1.2 & 5.8 & -28.8 & 68.8 & 5.0 & 13.0 & 25.7 \\
Qwen3.6-27B & C1 & 0.0 & 2.8 & 8.2 & -15.7 & 78.3 & 8.3 & 22.0 & 39.3 \\
Qwen3.6-27B & C2 & 11.3 & 2.5 & 3.0 & -3.8 & 91.5 & -2.3 & 10.0 & 14.0 \\
Qwen3.6-27B & C3 & 14.0 & 9.8 & 14.8 & -2.7 & 92.0 & -7.7 & 17.7 & 20.0 \\
Qwen3.6-27B & C4 & 67.0 & 12.8 & 21.6 & -1.8 & 91.8 & -2.0 & 19.3 & 40.0 \\
\bottomrule
\end{tabular}
\caption{Complete selected cue profiles on Core 300 (signed pp). Verified accuracy and valid-response rate are 100\% in every cell. C0--C4 are Base, Sham, Flat, Ledger, and +Rule.}
\label{tab:intervention-core-profile}
\end{table*}
\begin{table*}[t]
\centering
\footnotesize
\setlength{\tabcolsep}{4pt}
\begin{tabular}{llrr}
\toprule
Model & Comparison & $\Delta$ history sensitivity & $\Delta |$order effect$|$ \\
\midrule
Qwen3.5-9B & Ledger $-$ Flat & $+0.0\;[0.0, 0.0]$ & $-0.5\;[-4.0, 2.8]$ \\
Qwen3.5-9B & Ledger $-$ Sham & $-0.3\;[-1.0, 0.0]$ & $-10.7\;[-14.7, -6.5]$$^{*}$ \\
Qwen3.5-9B & +Rule $-$ Ledger & $+0.3\;[0.0, 1.0]$ & $-0.7\;[-4.2, 3.0]$ \\
Qwen3.5-9B & Flat $-$ Base & $+0.0\;[0.0, 0.0]$ & $+12.0\;[7.8, 16.2]$$^{*}$ \\
\midrule
Qwen3.5-27B & Ledger $-$ Flat & $-0.3\;[-3.7, 2.7]$ & $-4.8\;[-7.5, -2.2]$$^{*}$ \\
Qwen3.5-27B & Ledger $-$ Sham & $+23.3\;[18.3, 28.0]$$^{*}$ & $-13.2\;[-16.7, -9.7]$$^{*}$ \\
Qwen3.5-27B & +Rule $-$ Ledger & $+41.7\;[36.0, 47.0]$$^{*}$ & $+0.2\;[-2.8, 3.2]$ \\
Qwen3.5-27B & Flat $-$ Base & $+23.3\;[18.7, 28.3]$$^{*}$ & $-36.3\;[-40.2, -32.3]$$^{*}$ \\
\midrule
Qwen3.6-27B & Ledger $-$ Flat & $+2.7\;[-1.3, 6.7]$ & $-1.2\;[-3.0, 0.5]$ \\
Qwen3.6-27B & Ledger $-$ Sham & $+14.0\;[10.0, 18.0]$$^{*}$ & $-13.0\;[-16.0, -10.3]$$^{*}$ \\
Qwen3.6-27B & +Rule $-$ Ledger & $+53.0\;[47.3, 58.7]$$^{*}$ & $-0.8\;[-2.3, 0.7]$ \\
Qwen3.6-27B & Flat $-$ Base & $+7.0\;[3.3, 10.7]$$^{*}$ & $-25.0\;[-28.5, -21.7]$$^{*}$ \\
\bottomrule
\end{tabular}
\caption{Core counterparts of the pre-specified intervention comparisons (pp; paired 95\% intervals). Stars use the same 48-test Holm family as the main table.}
\label{tab:intervention-core-contrasts}
\end{table*}
\begin{table*}[t]
\centering
\footnotesize
\setlength{\tabcolsep}{4pt}
\begin{tabular}{llrrr}
\toprule
Model & Comparison & $\Delta |$order: early$|$ & $\Delta |$order: late$|$ & $\Delta$ time \\
\midrule
Qwen3.5-9B & Ledger $-$ Flat & $-2.4\;[-6.4, 1.4]$ & $+2.6\;[-1.0, 6.2]$ & $+0.7\;[-2.2, 3.4]$ \\
Qwen3.5-9B & Ledger $-$ Sham & $-9.4\;[-14.2, -4.6]$ & $-7.4\;[-11.2, -3.6]$ & $+10.8\;[7.5, 13.9]$ \\
Qwen3.5-9B & +Rule $-$ Ledger & $+1.8\;[-2.8, 6.4]$ & $-2.6\;[-6.0, 0.6]$ & $-9.2\;[-12.2, -6.1]$ \\
Qwen3.5-9B & Flat $-$ Base & $+12.8\;[8.0, 17.4]$ & $+6.6\;[2.8, 10.6]$ & $+11.1\;[7.7, 14.5]$ \\
\midrule
Qwen3.5-27B & Ledger $-$ Flat & $-3.0\;[-5.8, -0.2]$ & $-3.2\;[-6.8, 1.4]$ & $-0.5\;[-2.7, 1.6]$ \\
Qwen3.5-27B & Ledger $-$ Sham & $-7.6\;[-11.0, -4.4]$ & $-17.4\;[-22.0, -12.2]$ & $+23.1\;[20.2, 26.1]$ \\
Qwen3.5-27B & +Rule $-$ Ledger & $-2.0\;[-5.4, 1.8]$ & $-1.2\;[-3.2, 1.0]$ & $-11.8\;[-14.8, -8.8]$ \\
Qwen3.5-27B & Flat $-$ Base & $-37.0\;[-41.6, -32.4]$ & $-33.4\;[-37.8, -29.0]$ & $+30.8\;[27.7, 33.7]$ \\
\midrule
Qwen3.6-27B & Ledger $-$ Flat & $-0.6\;[-3.2, 2.0]$ & $-1.2\;[-2.4, 1.6]$ & $-0.3\;[-1.7, 1.1]$ \\
Qwen3.6-27B & Ledger $-$ Sham & $-14.4\;[-18.4, -10.4]$ & $-4.2\;[-6.4, -0.6]$ & $+13.1\;[10.8, 15.5]$ \\
Qwen3.6-27B & +Rule $-$ Ledger & $-0.8\;[-2.8, 1.2]$ & $+0.8\;[-1.2, 1.6]$ & $+0.6\;[-0.8, 2.0]$ \\
Qwen3.6-27B & Flat $-$ Base & $-21.8\;[-25.8, -18.0]$ & $-25.0\;[-28.8, -21.2]$ & $+22.6\;[19.9, 25.3]$ \\
\bottomrule
\end{tabular}
\caption{Pointwise follow-up contrasts for conditional order and event-time sensitivity (pp; paired 95\% intervals). These separate the two temporal contexts behind the aggregate order endpoint.}
\label{tab:intervention-order}
\end{table*}
\begin{table*}[t]
\centering
\footnotesize
\setlength{\tabcolsep}{4pt}
\begin{tabular}{llrrr}
\toprule
Model & Comparison & $\Delta$ lineage & $\Delta$ repetition & $\Delta$ speakers \\
\midrule
Qwen3.5-9B & Ledger $-$ Flat & $+21.8\;[16.6, 26.6]$ & $-16.0\;[-20.8, -11.2]$ & $-0.4\;[-4.6, 3.8]$ \\
Qwen3.5-9B & Ledger $-$ Sham & $+6.2\;[0.8, 11.6]$ & $-2.0\;[-6.4, 2.8]$ & $+0.0\;[-4.2, 4.2]$ \\
Qwen3.5-9B & +Rule $-$ Ledger & $+22.0\;[17.2, 26.8]$ & $-0.4\;[-4.6, 3.6]$ & $-3.6\;[-7.4, 0.0]$ \\
Qwen3.5-9B & Flat $-$ Base & $-20.6\;[-25.6, -15.8]$ & $+11.0\;[6.4, 15.8]$ & $+4.2\;[0.6, 7.8]$ \\
\midrule
Qwen3.5-27B & Ledger $-$ Flat & $+9.4\;[5.6, 13.2]$ & $+14.0\;[9.2, 18.8]$ & $+2.0\;[-1.6, 5.8]$ \\
Qwen3.5-27B & Ledger $-$ Sham & $-11.4\;[-16.4, -6.2]$ & $-16.2\;[-20.2, -12.6]$ & $+7.2\;[3.8, 10.6]$ \\
Qwen3.5-27B & +Rule $-$ Ledger & $+19.2\;[15.4, 23.0]$ & $-5.0\;[-8.8, -1.0]$ & $+1.0\;[-2.6, 4.6]$ \\
Qwen3.5-27B & Flat $-$ Base & $-14.6\;[-18.6, -10.8]$ & $-27.4\;[-32.6, -22.2]$ & $+4.0\;[0.4, 7.6]$ \\
\midrule
Qwen3.6-27B & Ledger $-$ Flat & $+10.6\;[7.4, 14.0]$ & $-6.6\;[-11.0, -2.0]$ & $+6.2\;[2.2, 10.2]$ \\
Qwen3.6-27B & Ledger $-$ Sham & $-17.2\;[-21.4, -13.0]$ & $-16.4\;[-21.2, -11.6]$ & $-4.4\;[-9.0, 0.4]$ \\
Qwen3.6-27B & +Rule $-$ Ledger & $+19.4\;[15.8, 23.0]$ & $+7.6\;[3.0, 12.2]$ & $+0.8\;[-3.8, 5.2]$ \\
Qwen3.6-27B & Flat $-$ Base & $-13.8\;[-17.2, -10.4]$ & $-5.8\;[-10.4, -1.2]$ & $-5.2\;[-9.2, -1.2]$ \\
\bottomrule
\end{tabular}
\caption{Pointwise intervention contrasts for evidence lineage, repeated reports, and speaker multiplicity (pp; paired 95\% intervals).}
\label{tab:intervention-secondary}
\end{table*}
\begin{table*}[t]
\centering
\footnotesize
\setlength{\tabcolsep}{4pt}
\begin{tabular}{lrrrr}
\toprule
Model & $\Delta$ history & $\Delta |$order$|$ & $\Delta$ time & $\Delta$ lineage \\
\midrule
Qwen3.5-9B & $+0.2\;[-0.4, 1.0]$ & $+18.2\;[14.9, 21.4]$ & $+1.0\;[-2.1, 4.0]$ & $-5.0\;[-9.6, -0.4]$ \\
Qwen3.5-27B & $-0.8\;[-1.6, -0.2]$ & $-25.8\;[-28.8, -22.9]$ & $+7.2\;[3.8, 10.5]$ & $+6.2\;[1.0, 11.4]$ \\
Qwen3.6-27B & $-3.8\;[-5.6, -2.0]$ & $-15.0\;[-17.8, -12.2]$ & $+9.2\;[6.5, 12.0]$ & $+14.0\;[10.0, 18.2]$ \\
\bottomrule
\end{tabular}
\caption{Sham minus Base: context changes without epistemic fact restatement (pp; pointwise 95\% paired intervals).}
\label{tab:sham}
\end{table*}
\begin{table}[t]
\centering
\resizebox{0.5\textwidth}{!}{
\setlength{\tabcolsep}{4pt}
\begin{tabular}{lrrrr}
\toprule
Model & Comparable & Both valid & Changed & Changed (\%) \\
\midrule
Qwen3.5-9B & 7520 & 7520 & 105 & 1.40 \\
Qwen3.5-27B & 7520 & 7520 & 30 & 0.40 \\
Qwen3.6-27B & 7520 & 7505 & 29 & 0.39 \\
\bottomrule
\end{tabular}}
\caption{Rerun Base consistency on exactly comparable original requests. The intervention contrasts use the new within-batch baseline, not a mixture of the two runs.}
\label{tab:baseline-rerun}
\end{table}

\paragraph{Protective outcome.}
All 1,000 verified decisions per model and intervention condition are correct and format-valid. This is an observed 100\% inclusive and valid-pair accuracy across all 15 model--condition cells. The intervention's pre-set descriptive preservation margin was a decrease of no more than two percentage points.

\paragraph{Material scope.}
The intervention expands the earlier 300-world history and 240-world policy modules to all 500 worlds through the same diagnostic overlays. The factorial and verification tasks preserve each scenario's original evidence graph; the history and policy tasks use the fixed two-root diagnostic constructions. This is why the intervention has its own baseline and denominator.

\subsection{Additional Coverage and Robustness Tables}
\label{app:robustness}
Table~\ref{tab:paired-coverage} reports the actual contributing Core worlds behind the principal paired estimates. Relation-level history-swap estimates in Tables~\ref{tab:relation-core}--\ref{tab:relation-all} are descriptive subdivisions of the same benchmark. The diagnostic and intervention cluster checks retain their own original resampling units rather than pool requests as independent observations.
\begin{table*}[t]
\centering
\footnotesize
\setlength{\tabcolsep}{4pt}
\begin{tabular}{lrrrr}
\toprule
Model & Swap $n$ & Eight-condition $n$ & Text/time $n$ & Swap missing bounds \\
\midrule
Qwen3.5-2B & 0 & 0 & 0 & $[-100.0, 100.0]$ \\
Qwen3.5-4B & 139 & 50 & 105 & $[-41.7, 49.3]$ \\
Qwen3.5-9B & 300 & 300 & 300 & $[0.3, 0.3]$ \\
Qwen3.5-27B & 300 & 300 & 300 & $[0.7, 0.7]$ \\
Qwen3.6-27B & 300 & 294 & 297 & $[4.0, 4.0]$ \\
Qwen3.8-27B & 300 & 300 & 300 & $[0.0, 0.0]$ \\
InternVL3.5-8B & 300 & 300 & 300 & $[4.7, 4.7]$ \\
InternVL3.5-14B & 256 & 185 & 250 & $[-7.3, 13.0]$ \\
Grok 4.3 & 300 & 300 & 300 & $[73.0, 73.0]$ \\
Claude Opus 4.7 & 300 & 289 & 300 & $[33.0, 33.0]$ \\
GPT-5.4 & 300 & 300 & 300 & $[6.3, 6.3]$ \\
GPT-6 Sol & 300 & 300 & 300 & $[100.0, 100.0]$ \\
\bottomrule
\end{tabular}
\caption{Core paired-world coverage (300 planned worlds) and the full-plan worst-case history-swap bounds (pp). Accuracy denominators are planned requests and are reported separately.}
\label{tab:paired-coverage}
\end{table*}
\begin{table*}[t]
\centering
\footnotesize
\setlength{\tabcolsep}{4pt}
\begin{tabular}{lrrrrrrrrr}
\toprule
Relation & $n$ & 3.5-9B & 3.5-27B & 3.6-27B & 3.8-27B & Grok & Claude & GPT-5.4 & GPT-6 \\
\midrule
Assignment & 24 & 4.2 & 4.2 & 12.5 & 0.0 & 79.2 & 41.7 & 4.2 & 100.0 \\
Custody & 24 & 0.0 & 0.0 & 8.3 & 0.0 & 66.7 & 29.2 & 8.3 & 100.0 \\
Location & 27 & 0.0 & 0.0 & 0.0 & 0.0 & 81.5 & 51.9 & 11.1 & 100.0 \\
Membership & 20 & 0.0 & 0.0 & 0.0 & 0.0 & 65.0 & 40.0 & 10.0 & 100.0 \\
Outcome & 27 & 0.0 & 3.7 & 0.0 & 0.0 & 66.7 & 25.9 & 11.1 & 100.0 \\
Permission & 24 & 0.0 & 0.0 & 0.0 & 0.0 & 79.2 & 25.0 & 0.0 & 100.0 \\
Priority & 24 & 0.0 & 0.0 & 8.3 & 0.0 & 75.0 & 45.8 & 0.0 & 100.0 \\
Quantity & 24 & 0.0 & 0.0 & 4.2 & 0.0 & 79.2 & 25.0 & 8.3 & 100.0 \\
Routing & 31 & 0.0 & 0.0 & 9.7 & 0.0 & 87.1 & 35.5 & 12.9 & 100.0 \\
Schedule & 24 & 0.0 & 0.0 & 0.0 & 0.0 & 58.3 & 29.2 & 4.2 & 100.0 \\
State & 27 & 0.0 & 0.0 & 3.7 & 0.0 & 70.4 & 25.9 & 3.7 & 100.0 \\
Version & 24 & 0.0 & 0.0 & 0.0 & 0.0 & 62.5 & 20.8 & 0.0 & 100.0 \\
\bottomrule
\end{tabular}
\caption{History-swap effects by relation on Core (pp). The first four systems are Qwen checkpoints. These strata describe where each system changes its choice under the same manipulation.}
\label{tab:relation-core}
\end{table*}
\begin{table*}[t]
\centering
\footnotesize
\setlength{\tabcolsep}{4pt}
\begin{tabular}{lrrrrrrrrr}
\toprule
Relation & $n$ & 3.5-9B & 3.5-27B & 3.6-27B & 3.8-27B & Grok & Claude & GPT-5.4 & GPT-6 \\
\midrule
Assignment & 40 & 2.5 & 5.0 & 7.5 & 0.0 & 82.5 & 40.0 & 10.0 & 100.0 \\
Custody & 40 & 0.0 & 0.0 & 10.0 & 0.0 & 72.5 & 35.0 & 7.5 & 100.0 \\
Location & 45 & 0.0 & 2.2 & 0.0 & 0.0 & 75.6 & 44.4 & 8.9 & 100.0 \\
Membership & 36 & 0.0 & 0.0 & 2.8 & 0.0 & 72.2 & 38.9 & 11.1 & 100.0 \\
Outcome & 45 & 4.4 & 4.4 & 2.2 & 0.0 & 71.1 & 24.4 & 11.1 & 100.0 \\
Permission & 40 & 0.0 & 0.0 & 2.5 & 0.0 & 77.5 & 20.0 & 2.5 & 100.0 \\
Priority & 40 & 0.0 & 0.0 & 5.0 & 0.0 & 77.5 & 37.5 & 2.5 & 100.0 \\
Quantity & 40 & 0.0 & 0.0 & 2.5 & 0.0 & 75.0 & 25.0 & 10.0 & 100.0 \\
Routing & 49 & 0.0 & 0.0 & 12.2 & 0.0 & 85.7 & 39.6 & 8.2 & 100.0 \\
Schedule & 40 & 0.0 & 0.0 & 0.0 & 0.0 & 57.5 & 35.0 & 10.0 & 100.0 \\
State & 45 & 0.0 & 2.2 & 2.2 & 0.0 & 71.1 & 22.2 & 6.7 & 100.0 \\
Version & 40 & 0.0 & 0.0 & 2.5 & 0.0 & 70.0 & 22.5 & 5.0 & 100.0 \\
\bottomrule
\end{tabular}
\caption{History-swap effects by relation on all 500 scenarios (pp). The first four systems are Qwen checkpoints. These strata describe where each system changes its choice under the same manipulation.}
\label{tab:relation-all}
\end{table*}
\begin{table*}[t]
\centering
\footnotesize
\setlength{\tabcolsep}{4pt}
\begin{tabular}{lrrr}
\toprule
Model & Original history & Sham appended & Counts appended \\
\midrule
Qwen3.5-2B & $\text{--}$ & $\text{--}$ & $\text{--}$ \\
Qwen3.5-4B & $9.4\;[5.0, 14.4]$ & $10.7\;[6.7, 14.7]$ & $6.6\;[3.3, 10.0]$ \\
Qwen3.5-9B & $0.7\;[0.0, 1.7]$ & $0.3\;[0.0, 1.0]$ & $0.0\;[0.0, 0.0]$ \\
Qwen3.5-27B & $0.3\;[0.0, 1.0]$ & $1.0\;[0.0, 2.3]$ & $21.7\;[17.3, 26.3]$ \\
Qwen3.6-27B & $4.7\;[2.7, 7.0]$ & $9.0\;[5.7, 12.3]$ & $23.0\;[18.3, 27.7]$ \\
Qwen3.8-27B & $0.0\;[0.0, 0.0]$ & $0.0\;[0.0, 0.0]$ & $0.0\;[0.0, 0.0]$ \\
\bottomrule
\end{tabular}
\caption{Dedicated diagnostic history-swap effects when original histories are unchanged, a sham is appended, or exact historical counts are appended (pp; supplied 95\% intervals). These 300-world diagnostic conditions precede the separate 500-world intervention and are not substitutes for its rerun baseline.}
\label{tab:count-append}
\end{table*}
\begin{table*}[t]
\centering
\footnotesize
\setlength{\tabcolsep}{4pt}
\begin{tabular}{lrrrr}
\toprule
Model & World resampling & Renderer resampling & Event-graph resampling & Holm $p$ \\
\midrule
Qwen3.5-2B & $\text{--}$ & \textemdash & \textemdash & \textemdash \\
Qwen3.5-4B & $2.0\;[-1.0, 4.0]$ & $2.0\;[-2.2, 7.4]$ & $2.0\;[-1.9, 8.0]$ & 1.00 \\
Qwen3.5-9B & $3.0\;[-1.8, 7.7]$ & $3.0\;[-1.2, 7.6]$ & $3.0\;[-2.0, 8.2]$ & 1.00 \\
Qwen3.5-27B & $-1.3\;[-3.7, 1.0]$ & $-1.3\;[-3.6, 1.1]$ & $-1.3\;[-4.0, 1.2]$ & 1.00 \\
Qwen3.6-27B & $-0.2\;[-2.2, 1.9]$ & $-0.2\;[-2.4, 2.0]$ & $-0.2\;[-1.6, 1.4]$ & 1.00 \\
Qwen3.8-27B & $0.5\;[-0.3, 1.5]$ & $0.5\;[-0.3, 1.5]$ & $0.5\;[0.0, 1.2]$ & 1.00 \\
\bottomrule
\end{tabular}
\caption{Pre-specified six-Qwen Core $Q	imes I$ result and stimulus-cluster sensitivity in the dedicated diagnostic study (pp). The estimable primary tests all have adjusted $p=1.0$.}
\label{tab:qi-clusters}
\end{table*}
\begin{table*}[t]
\centering
\footnotesize
\setlength{\tabcolsep}{4pt}
\begin{tabular}{llrr}
\toprule
Model & Comparison & $\Delta$ history & $\Delta |$order$|$ \\
\midrule
Qwen3.5-9B & Ledger $-$ Flat & $+0.0\;[-0.6, 0.6]$ & $+0.1\;[-2.9, 3.0]$ \\
Qwen3.5-9B & Ledger $-$ Sham & $-0.4\;[-1.2, 0.4]$ & $-8.4\;[-12.0, -5.0]$ \\
Qwen3.5-9B & +Rule $-$ Ledger & $+0.4\;[0.0, 1.0]$ & $-0.4\;[-3.3, 2.6]$ \\
Qwen3.5-9B & Flat $-$ Base & $-0.2\;[-0.6, 0.0]$ & $+9.7\;[6.7, 12.7]$ \\
\midrule
Qwen3.5-27B & Ledger $-$ Flat & $-0.4\;[-3.0, 2.4]$ & $-4.5\;[-6.5, -2.5]$ \\
Qwen3.5-27B & Ledger $-$ Sham & $+25.0\;[20.6, 29.5]$ & $-13.9\;[-16.9, -11.0]$ \\
Qwen3.5-27B & +Rule $-$ Ledger & $+39.2\;[33.9, 44.5]$ & $-0.4\;[-2.4, 2.0]$ \\
Qwen3.5-27B & Flat $-$ Base & $+24.6\;[20.4, 28.7]$ & $-35.2\;[-38.2, -32.1]$ \\
\midrule
Qwen3.6-27B & Ledger $-$ Flat & $+4.2\;[1.0, 7.3]$ & $-1.1\;[-2.6, 0.4]$ \\
Qwen3.6-27B & Ledger $-$ Sham & $+17.0\;[13.6, 20.5]$ & $-9.5\;[-11.9, -7.1]$ \\
Qwen3.6-27B & +Rule $-$ Ledger & $+47.2\;[42.3, 52.1]$ & $-0.8\;[-2.2, 0.6]$ \\
Qwen3.6-27B & Flat $-$ Base & $+9.0\;[5.7, 12.4]$ & $-23.4\;[-26.6, -20.3]$ \\
\bottomrule
\end{tabular}
\caption{Intervention main contrasts resampled by renderer rather than individual world (pp; supplied 95\% intervals). There are 120 renderer clusters in the All-500 analysis.}
\label{tab:intervention-renderer}
\end{table*}
\begin{table*}[t]
\centering
\footnotesize
\setlength{\tabcolsep}{4pt}
\begin{tabular}{llrrrrrrrr}
\toprule
Model & Cond. & Hist. & Q & I & Text & Time & Rep. & Spk. & Root \\
\midrule
Qwen3.5-9B & C0 & 0.8 & 11.5 & 18.5 & 12.9 & 32.9 & -2.5 & 6.7 & 38.3 \\
Qwen3.5-9B & C1 & 0.8 & 13.8 & 20.4 & 34.2 & 35.0 & -10.8 & 12.5 & 30.0 \\
Qwen3.5-9B & C2 & 0.8 & 9.0 & 21.5 & 25.0 & 50.0 & 0.8 & 20.8 & 12.5 \\
Qwen3.5-9B & C3 & 0.0 & 22.1 & 26.7 & 25.4 & 50.4 & -7.5 & 12.5 & 39.2 \\
Qwen3.5-9B & C4 & 0.8 & 20.6 & 36.5 & 25.8 & 37.5 & -11.7 & 6.7 & 62.5 \\
\midrule
Qwen3.5-27B & C0 & 0.8 & 2.9 & 15.0 & -42.5 & 56.7 & -2.5 & 7.5 & 62.5 \\
Qwen3.5-27B & C1 & 0.0 & 4.2 & 12.1 & -18.8 & 62.9 & 5.8 & 4.2 & 67.5 \\
Qwen3.5-27B & C2 & 22.5 & 4.8 & 10.6 & -8.8 & 85.4 & -25.8 & 11.7 & 44.2 \\
Qwen3.5-27B & C3 & 23.3 & 11.9 & 16.9 & -2.1 & 86.2 & -6.7 & 13.3 & 48.3 \\
Qwen3.5-27B & C4 & 63.3 & 14.4 & 26.5 & -1.7 & 72.5 & -18.3 & 15.0 & 70.0 \\
\midrule
Qwen3.6-27B & C0 & 5.8 & 2.3 & 6.5 & -30.0 & 67.5 & 1.7 & 15.8 & 35.8 \\
Qwen3.6-27B & C1 & 0.8 & 2.3 & 6.0 & -10.8 & 80.8 & 7.5 & 16.7 & 52.5 \\
Qwen3.6-27B & C2 & 15.8 & 2.9 & 5.4 & -5.0 & 92.5 & -1.7 & 9.2 & 19.2 \\
Qwen3.6-27B & C3 & 15.8 & 9.6 & 15.0 & -3.3 & 93.3 & -6.7 & 16.7 & 28.3 \\
Qwen3.6-27B & C4 & 62.5 & 12.9 & 25.0 & -2.5 & 94.2 & 0.8 & 15.8 & 49.2 \\
\bottomrule
\end{tabular}
\caption{Intervention profiles on Structural 120 (signed pp). This is a predefined material subset, not a paired manipulation of linguistic form.}
\label{tab:int-structural}
\end{table*}
\begin{table*}[t]
\centering
\footnotesize
\setlength{\tabcolsep}{4pt}
\begin{tabular}{llrrrrrrrr}
\toprule
Model & Cond. & Hist. & Q & I & Text & Time & Rep. & Spk. & Root \\
\midrule
Qwen3.5-9B & C0 & 1.2 & 3.4 & 14.1 & 6.2 & 33.8 & -6.2 & 15.0 & 32.5 \\
Qwen3.5-9B & C1 & 1.2 & 2.5 & 11.9 & 5.0 & 25.0 & 3.8 & 20.0 & 23.8 \\
Qwen3.5-9B & C2 & 0.0 & 2.5 & 3.1 & 3.8 & 23.8 & 11.2 & 7.5 & 13.8 \\
Qwen3.5-9B & C3 & 1.2 & 5.9 & 6.6 & 5.6 & 38.1 & -5.0 & 10.0 & 37.5 \\
Qwen3.5-9B & C4 & 1.2 & 9.7 & 12.8 & 5.0 & 21.2 & -1.2 & 12.5 & 43.8 \\
\midrule
Qwen3.5-27B & C0 & 3.8 & 3.1 & 4.4 & -31.2 & 45.0 & 3.8 & 8.8 & 57.5 \\
Qwen3.5-27B & C1 & 1.2 & 2.8 & 2.8 & -10.6 & 44.4 & 2.5 & 12.5 & 56.2 \\
Qwen3.5-27B & C2 & 37.5 & 3.8 & 2.5 & 1.9 & 71.9 & -16.2 & 10.0 & 46.2 \\
Qwen3.5-27B & C3 & 35.0 & 9.4 & 9.4 & 1.9 & 68.1 & -11.2 & 13.8 & 61.3 \\
Qwen3.5-27B & C4 & 63.7 & 20.0 & 24.4 & 4.4 & 66.9 & -2.5 & 6.2 & 80.0 \\
\midrule
Qwen3.6-27B & C0 & 3.8 & 4.4 & 6.9 & -13.8 & 65.0 & -1.2 & 25.0 & 28.7 \\
Qwen3.6-27B & C1 & 3.8 & 4.4 & 5.6 & 1.9 & 66.9 & 2.5 & 23.8 & 40.0 \\
Qwen3.6-27B & C2 & 18.8 & 4.1 & 4.7 & 1.2 & 83.8 & -5.0 & 13.8 & 11.2 \\
Qwen3.6-27B & C3 & 35.0 & 14.1 & 18.4 & 1.2 & 78.8 & -18.8 & 12.5 & 41.2 \\
Qwen3.6-27B & C4 & 61.3 & 20.3 & 31.6 & 1.9 & 81.9 & -3.8 & 12.5 & 56.2 \\
\bottomrule
\end{tabular}
\caption{Intervention profiles on Naturalistic 80 (signed pp). This is a predefined material subset, not a paired manipulation of linguistic form.}
\label{tab:int-natural}
\end{table*}

\subsection{Coverage of Supplementary Results}
\label{app:coverage-map}
The main text reports the cross-system profile, identification--influence contrast, source and temporal controls, and the pre-specified intervention effects. The appendices retain equal-history judgments, exact-count reconstruction, verified conditional denominators, null factorial findings, predictive checks, sham changes, conditional order contrasts, and baseline rerun consistency. Relation-, renderer-, and event-graph-specific estimates remain in the supplied result archives and in the accompanying source-disposition index; individual numerical exports are not typeset as thousands of redundant table rows. Incomplete runs, duplicate summaries, and non-estimable quantities are not promoted to additional experiments. The old 50-world pilot and 24 G1-X development scenarios are excluded from all formal benchmark counts.

% =========================================================
% Appendix A: Dataset Statistics
% =========================================================

\newpage

\section{Dataset Statistics}
\label{app:dataset-statistics}
\begin{table}[t]
\centering
\small
\setlength{\tabcolsep}{5pt}
\begin{tabular}{lrr}
\toprule
Subset & Worlds & Fraction \\
\midrule
Controlled Core & 300 & 60\% \\
Structural Generalization & 120 & 24\% \\
Controlled Naturalized & 80 & 16\% \\
\midrule
\textbf{Total} & \textbf{500} & \textbf{100\%} \\
\bottomrule
\end{tabular}
\caption{Composition of the frozen 500-world evaluation set.}
\label{tab:dataset-subsets}
\end{table}

\subsection{Evaluation Set and Overall Scale}
\label{app:dataset-overview}

The frozen evaluation set contains 500 fictional worlds. These worlds are
evaluation strata rather than train, validation, and test splits. We divide
them into 300 Controlled Core worlds, 120 Structural Generalization worlds,
and 80 Controlled Naturalized worlds. The same frozen world bank is used
throughout the benchmark and subsequent diagnostic analyses.
The Controlled Core provides the most tightly matched experimental
comparisons. Structural Generalization changes event and provenance
structures while retaining the same target judgments and experimental
factors. Controlled Naturalized worlds preserve the symbolic variables and
gold records while expressing them through more naturalized discourse.
All model inputs are English text. References to photographs, logs,
documents, and audio recordings are textual descriptions of fictional
evidence; no image, audio, or video input is provided to the models.

\begin{table}[t]
\centering
\small
\setlength{\tabcolsep}{6pt}
\begin{tabular}{lr}
\toprule
Dataset statistic & Count \\
\midrule
Worlds / target entities & 500 \\
Relation families & 12 \\
Domains & 25 \\
Discourse formats & 10 \\
Evidence-quality realizations & 4 \\
Background-event structures & 10 \\
Labeled source-graph signatures & 22 \\
Structural-generalization axes & 5 \\
Requests per world & 48 \\
Requests per model & 24,000 \\
Evaluated models & 13 \\
Model--request cells & 312,000 \\
\bottomrule
\end{tabular}
\caption{Summary statistics for the benchmark. Source-graph signatures
retain node labels and should not be interpreted as graph-isomorphism
classes.}
\label{tab:dataset-summary}
\end{table}

\subsection{Relation Families and Domains}
\label{app:relations-domains}

The 500 worlds span 12 relation families. Table~\ref{tab:relation-dist}
reports their distribution across the three evaluation strata.
\begin{table}[t]
\centering
\small
\resizebox{0.5\textwidth}{!}{
\setlength{\tabcolsep}{7pt}
\begin{tabular}{lrrrr}
\toprule
Relation family & Core & Structural & Naturalized & Total \\
\midrule
Assignment & 24 & 10 & 6 & 40 \\
Custody & 24 & 10 & 6 & 40 \\
Location & 27 & 10 & 8 & 45 \\
Membership & 20 & 10 & 6 & 36 \\
Outcome & 27 & 10 & 8 & 45 \\
Permission & 24 & 10 & 6 & 40 \\
Priority & 24 & 10 & 6 & 40 \\
Quantity & 24 & 10 & 6 & 40 \\
Routing & 31 & 10 & 8 & 49 \\
Schedule & 24 & 10 & 6 & 40 \\
State / mode & 27 & 10 & 8 & 45 \\
Version & 24 & 10 & 6 & 40 \\
\midrule
\textbf{Total} & \textbf{300} & \textbf{120} & \textbf{80} & \textbf{500} \\
\bottomrule
\end{tabular}}
\caption{Distribution of relation families. Structural Generalization contains
10 worlds from each relation family.}
\label{tab:relation-dist}
\end{table}
The scenarios additionally cover 25 domains:
agriculture, community centers, construction, event planning, facilities,
fleet operations, homes, hotels, IT operations, libraries, makerspaces,
manufacturing, media production, offices, research laboratories,
restaurants, retail, schools, shipping, software development, sports clubs,
travel, universities, utilities, and warehouses. Each domain contributes
20 worlds to the full set.

\subsection{Discourse Formats}
\label{app:discourse-formats}

We vary the surface organization of the scenarios independently of their
underlying symbolic records. The ten discourse formats each occur in
50 worlds.

\begin{table}[ht]
\centering
\small
\begin{tabular}{lr}
\toprule
Discourse format & Worlds \\
\midrule
Email thread & 50 \\
First-person report & 50 \\
Group chat & 50 \\
Handoff notes & 50 \\
Incident report & 50 \\
Meeting minutes & 50 \\
Task-update thread & 50 \\
Third-person narrative & 50 \\
Timestamped log & 50 \\
Two-person dialogue & 50 \\
\midrule
\textbf{Total} & \textbf{500} \\
\bottomrule
\end{tabular}
\caption{Distribution of discourse formats.}
\label{tab:discourse-formats}
\end{table}

\subsection{Evidence-Quality Manipulation}
\label{app:evidence-quality}

Current evidence quality is manipulated through the identifiability of the
target in the current evidence record. Each world uses one of four textual
evidence realizations. In the high-quality condition, the complete target
identifier is available; in the limited-quality condition, part of the
identifier is unavailable while the reported state value remains visible.
Thus, the manipulation changes how clearly the current evidence can be
attributed to the target rather than directly declaring the report true or
false.
Each realization is assigned to 125 worlds.

\begin{table*}[t]
\centering
\footnotesize
\setlength{\tabcolsep}{5pt}
\begin{tabular}{p{2.2cm}p{6.1cm}p{6.1cm}}
\toprule
Realization & High quality & Limited quality \\
\midrule
Photo identifier &
The photograph shows the entire identifying tag and the state indicator. &
The photograph shows only the first part of the identifying tag; the state
indicator remains visible. \\

Log identifier &
The log extract retains the complete target identifier and the recorded
value. &
The log extract retains only the prefix of the target identifier; the
recorded value remains visible. \\

Document identifier &
The document copy shows the complete case identifier and its value field. &
The document copy masks the last part of the case identifier; its value
field remains visible. \\

Audio identifier &
The audio clip preserves the complete spoken target identifier and the
spoken value. &
The audio clip loses the end of the spoken target identifier; the spoken
value remains audible. \\
\bottomrule
\end{tabular}
\caption{The four textual realizations of the evidence-quality manipulation.
These are descriptions of fictional evidence, not multimodal model inputs.}
\label{tab:quality-realizations}
\end{table*}

\begin{table*}[t]
\centering
\small
\setlength{\tabcolsep}{7pt}
\begin{tabular}{lrrrl}
\toprule
Track & Req./world & Req./model & 13-model cells & Main content \\
\midrule
Representation & 28 & 14,000 & 182,000 &
P1/P2 and reliability-history probes \\
Factorial Integration & 8 & 4,000 & 52,000 &
$R\times Q\times I$ final choices \\
Reliability Utilization & 2 & 1,000 & 13,000 &
HL/LH reliability swap \\
Cue Competition & 8 & 4,000 & 52,000 &
Time/order and report-structure contrasts \\
Verified Reasoning & 2 & 1,000 & 13,000 &
Verified-A/B final judgments \\
\midrule
\textbf{Total} & \textbf{48} & \textbf{24,000} &
\textbf{312,000} & \\
\bottomrule
\end{tabular}
\caption{Request composition of the 13-model benchmark.}
\label{tab:benchmark-track-counts}
\end{table*}
\subsection{Background Events and Structural Generalization}
\label{app:structural-generalization}

The world bank contains ten background-event structures, including alias
bindings, conditional releases, historical returns, parallel entities,
planned-versus-effective changes, proposals that have not taken effect,
scope copies, single changes, supersession chains, and temporary
intermediate states. These events diversify the reasoning context without
providing a trusted verification of the target-time state.
Structural Generalization varies five structural properties. Each axis
contains 24 worlds, for a total of 120.

\begin{table}[t]
\centering
\small
\setlength{\tabcolsep}{5pt}
\begin{tabular}{lr}
\toprule
Structural axis & Worlds \\
\midrule
Source count & 24 \\
Evidence-graph topology & 24 \\
Dependency pattern & 24 \\
Relay-chain length & 24 \\
Role configuration & 24 \\
\midrule
\textbf{Total} & \textbf{120} \\
\bottomrule
\end{tabular}
\caption{Axes used in the Structural Generalization subset.}
\label{tab:structural-axes}
\end{table}

Source-count worlds contain either three or four original sources.
Evidence-graph-topology worlds vary the mapping from original records to
three or four final messages. Dependency-pattern worlds vary whether relay
paths share an archival intermediary. Relay-chain-length worlds contain two-
or three-layer faithful relay chains. Role-configuration worlds vary whether
observers relay their own records or cross-relay another observer's record.

\subsection{Benchmark Request Composition}
\label{app:benchmark-request-composition}

Each world gives rise to 48 benchmark requests. The five tracks and their
request counts are shown in Table~\ref{tab:benchmark-track-counts}.
Of the 24,000 requests per model, 15,000 have an objective target and 9,000
are unverified support-choice requests without a unique gold state. The
benchmark contains 20,000 multiple-choice JSON requests and 4,000
integer-count requests per model.

\subsection{Counterbalancing}
\label{app:counterbalancing}

Default candidate-option order is balanced across the world bank:
250 worlds use the AB ordering and 250 use BA. The identity of the source
providing the shared evidence root is also balanced: O1 provides the shared
root in 250 worlds and O2 in the remaining 250.
The Verified-A and Verified-B versions of a world share the same underlying
condition and differ only in the trusted target-time verification value.
Semantic state labels A/B are distinct from the letter labels used for
multiple-choice options; option order is counterbalanced independently.

% =========================================================
% Appendix B: Supplementary Prompts and Evaluation Protocol
% =========================================================

\newpage
\section{Supplementary Prompts and Evaluation Protocol}
\label{app:protocol1}

\subsection{Independent Context Branches}
\label{app:independent-contexts}

Every probe is evaluated in an independent context. A request contains one
system message and one user message consisting of the complete information
required for that condition followed by a single question. Answers,
reasoning traces, or decisions from earlier probes are never inserted into
later requests.

Thus, an identification probe and a final decision for the same world are
separate model calls rather than turns in a shared conversation. In
particular, the model does not see its answer to a reliability-identification
probe when answering the corresponding final-choice question.

The general request structure is:

\begin{small}
\begin{verbatim}
SYSTEM:
[output-format instruction]

USER:
[fictional setting, target attribute, candidate states]
[earlier verified source histories, when present]
[trusted initial record]
[current evidence records and relay relationships]
[event and receipt times]
[trusted target-time verification, verified branches only]

[one probe question]
[lettered options, or integer-output instruction]
\end{verbatim}
\end{small}

No few-shot answer demonstrations are used. Earlier source histories are part
of the fictional evidence to be evaluated, rather than examples of model
responses.

\subsection{System Prompts}
\label{app:system-prompts}

\paragraph{Multiple-choice probes.}
All multiple-choice probes use the following system instruction:

\begin{small}
\begin{verbatim}
Answer the question using only the supplied fictional case.
Return only a JSON object with one key, "answer", whose value
is the option letter. Do not include an explanation.
\end{verbatim}
\end{small}

\paragraph{Exact-count probes.}
P3a and P3b use:

\begin{small}
\begin{verbatim}
Answer the question using only the supplied fictional case.
Return exactly one integer from 0 to 8, with no JSON, words,
or explanation.
\end{verbatim}
\end{small}

The user message additionally ends with:

\begin{small}
\begin{verbatim}
Return one integer from 0 to 8.
\end{verbatim}
\end{small}

\subsection{Identification and Decision Probes}
\label{app:probe-prompts}

Below we show the frozen question forms using one representative location
world. Entity names, attributes, candidate values, and option order vary
according to the world.

\paragraph{P1: Initial-state identification.}

\begin{small}
\begin{verbatim}
What value did the trusted initial record give for the
storage location of reference binder LO-0001 at 10:00?

A. upper cabinet
B. lower cabinet
\end{verbatim}
\end{small}

\paragraph{P2: Provenance and evidence-quality identification.}

\begin{small}
\begin{verbatim}
Which option correctly matches the later messages to their
original records and describes the identifying information
available in those records?

A. message 1 cites e1; message 2 cites e2;
   identifying information is complete.
B. message 1 cites e1; message 2 cites e2;
   identifying information is partly unavailable.
C. message 1 cites e1; message 2 cites e1;
   identifying information is complete.
D. message 1 cites e1; message 2 cites e1;
   identifying information is partly unavailable.
\end{verbatim}
\end{small}

\paragraph{P3a/P3b: Exact source-history counts.}

\begin{small}
\begin{verbatim}
In the eight fully verified earlier cases for O1,
how many reports matched their trusted verification?

Return one integer from 0 to 8.
\end{verbatim}
\end{small}

P3b uses the same form for O2.

\paragraph{P3c: Relative source reliability.}

\begin{small}
\begin{verbatim}
Across the supplied earlier verification records, which
source had the higher observed proportion of correct reports?

A. O1 had the higher proportion.
B. O2 had the higher proportion.
C. Their proportions were equal.
D. There is insufficient verification information.
\end{verbatim}
\end{small}

\paragraph{P4: Unverified final support judgment.}

\begin{small}
\begin{verbatim}
At 14:00 today, which candidate value for the storage
location of reference binder LO-0001 is better supported
by the supplied evidence?

A. upper cabinet
B. lower cabinet
\end{verbatim}
\end{small}

P4 does not receive a unique gold label unless the world is in a verified
branch.

\subsection{Factorial Integration Conditions}
\label{app:factorial-prompts}

The main factorial design crosses historical source reliability ($R$),
current evidence quality ($Q$), and evidence-record ancestry ($I$).

\begin{table}[t]
\centering
\small
\begin{tabular}{llll}
\toprule
Condition & Reliability & Quality & Ancestry \\
\midrule
C1 & HH & High & Distinct \\
C2 & HH & High & Shared \\
C3 & HH & Limited & Distinct \\
C4 & HH & Limited & Shared \\
C5 & LL & High & Distinct \\
C6 & LL & High & Shared \\
C7 & LL & Limited & Distinct \\
C8 & LL & Limited & Shared \\
\bottomrule
\end{tabular}
\caption{The $2\times2\times2$ factorial conditions. HH means that each
source matches trusted verification in 7 of 8 earlier cases; LL means
4 of 8 for each source.}
\label{tab:factorial-conditions}
\end{table}

All eight conditions include a trusted initial state at 10:00 and later
evidence describing 12:00, received at 13:00. Ordinary P4 questions concern
14:00 and do not provide a trusted verification of that target time.

The $R$ manipulation changes only the earlier verified source histories.
The $Q$ manipulation changes the completeness of target identification in
the current evidence record. The $I$ manipulation changes whether current
reports point to distinct original evidence records or to a shared record.

\subsection{Reliability-Swap Conditions}
\label{app:reliability-swap}

The reliability diagnostic uses four source-history profiles:

\begin{table}[t]
\centering
\small
\begin{tabular}{lccc}
\toprule
Profile & O1 history & O2 history & P3c target \\
\midrule
HH & 7/8 & 7/8 & Equal \\
LL & 4/8 & 4/8 & Equal \\
HL & 7/8 & 4/8 & O1 \\
LH & 4/8 & 7/8 & O2 \\
\bottomrule
\end{tabular}
\caption{Source-history profiles used in the reliability diagnostics.}
\label{tab:history-profiles}
\end{table}

For the reliability-swap P4 contrast, O1 supports semantic state A and O2
supports semantic state B. Evidence quality is held at the limited level and
the two current evidence records have distinct roots. The HL/LH comparison
therefore changes which currently disagreeing source has the stronger
verified history while keeping the current claims fixed.

\subsection{Text Order and Event Time}
\label{app:text-time}

The temporal-control family independently varies presentation order and
reported event time.

\begin{table}[t]
\centering
\small
\begin{tabular}{ll}
\toprule
Factor & Conditions \\
\midrule
Text order & Initial block first / Reports block first \\
Report event time & 08:00 / 12:00 \\
\bottomrule
\end{tabular}
\caption{Orthogonal text-order and event-time manipulation. The trusted
initial event always concerns 10:00; reports are received at 13:00; the
query concerns 14:00.}
\label{tab:text-time-conditions}
\end{table}

\begin{table}[t]
\centering
\small
\begin{tabular}{lr}
\toprule
Probe & Requests / model \\
\midrule
P1 & 4,000 \\
P2 & 4,000 \\
P3a & 2,000 \\
P3b & 2,000 \\
P3c & 2,000 \\
P4 & 9,000 \\
D4 & 1,000 \\
\midrule
\textbf{Total} & \textbf{24,000} \\
\bottomrule
\end{tabular}
\caption{Probe-level request counts for the final benchmark.}
\label{tab:probe-counts}
\end{table}

Changing paragraph order does not change the event or receipt timestamps.
A report describing 08:00 is therefore temporally earlier than the trusted
10:00 initial record even when the report paragraph appears later in the
prompt.

\subsection{Report Multiplicity and Evidence Lineage}
\label{app:report-structure}

The report-structure control separates the number of messages, the number of
relays, and the number of original evidence roots.

\begin{table*}[h]
\centering
\small
\begin{tabular}{clrrr}
\toprule
Condition & Report structure & Messages & Relays & Roots \\
\midrule
S1 & M1 relays e1 once & 1 & 1 & 1 \\
S2 & M1 repeats e1 twice & 2 & 1 & 1 \\
S3 & M1 and M2 both relay e1 & 2 & 2 & 1 \\
S4 & M1 relays e1; M2 relays independent e2 & 2 & 2 & 2 \\
\bottomrule
\end{tabular}
\caption{Report-structure conditions used in the benchmark. S2--S1
measures repetition, S3--S2 changes relay multiplicity, and S4--S3 changes
the number of independent evidence roots while holding two messages and
two relays fixed.}
\label{tab:report-structure}
\end{table*}

\begin{table*}[h]
\centering
\small
\begin{tabular}{lll}
\toprule
Arm & Name & Added material \\
\midrule
C0 & Baseline & No appendix \\
C1 & Matched sham & Source labels and length-matched organizational text \\
C2 & Flat restatement & Deterministically extracted epistemic facts in one paragraph \\
C3 & Epistemic ledger & The same facts grouped by source, record, message, and lineage \\
C4 & Ledger + rule & Full C3 ledger followed by a fixed weighting principle \\
\bottomrule
\end{tabular}
\caption{Prompt conditions used in the intervention study.}
\label{tab:intervention-arms}
\end{table*}
\subsection{Verified Reasoning}
\label{app:verified-prompt}

Verified-A and Verified-B branches append a separate trusted record that
directly establishes the target state at 14:00:

\begin{small}
\begin{verbatim}
A separate trusted verification was completed and received
at 14:00 today. It directly verifies that at 14:00 the
{target_relation} of {target_entity} was {verified_value}.
This is a trusted verification of the target time, not a
source report or a repetition of an earlier record.
\end{verbatim}
\end{small}

The two versions differ only in the verified state value. These branches
therefore have an objective gold answer and are scored by accuracy.

\subsection{Probe-Level Request Counts}
\label{app:probe-counts}

For completeness, Table~\ref{tab:probe-counts} reports the number of each
probe type in the 24,000-request benchmark for one model.

\subsection{Output Parsing}
\label{app:output-parsing}

Multiple-choice answers are considered format-valid only when the response is
a JSON object containing exactly one key, \texttt{answer}, whose value is a
legal option letter. Leading or trailing whitespace and one complete
\texttt{json} Markdown code fence are permitted. Extra fields, repeated keys,
explanatory text, or invalid option letters are treated as invalid format.

Exact-count probes are valid only when the stripped response contains one
integer character from 0 to 8. Invalid-format responses are not converted
into either candidate state in unverified P4 analyses.

% ---------------------------------------------------------
% Optional: include this subsection if Intervention is in the paper
% ---------------------------------------------------------

\subsection{Intervention Prompt Additions}
\label{app:intervention-prompts}

The intervention study preserves the original system instruction, scenario,
final question, and option order, and inserts one fixed appendix immediately
before the final question. Five conditions are used:
C2 and C3 contain the same fact set and differ in organization. C4 retains
the C3 ledger and appends the same general weighting principle to every
world, task, and model:

\begin{small}
\begin{verbatim}
When evaluating conflicting information, consider the
historical reliability of each source, the quality of the
current evidence, whether multiple reports derive from
independent evidence, and the time at which the reported
events occurred. Repeated reports that originate from the
same underlying evidence should not be treated as independent
corroboration. Textual presentation order alone should not
determine which claim is better supported.
\end{verbatim}
\end{small}

The rule contains no candidate-state answer and does not recommend selecting
a particular option.

\end{document}